\documentclass[aps,prd,preprint,superscriptaddress]{revtex4-2}
\usepackage{graphicx} % Required for inserting images
\usepackage{amsmath}
\usepackage{physics}
\usepackage{xcolor}
\usepackage{tikz}
\usepackage{soul}
\usetikzlibrary{decorations.pathmorphing}

\begin{document}
\title{Stationary Solution to Charged Hairy Black Hole in AdS$_4$: Kasner Interior, Rotating Shock Waves, and Fast Scrambling}
\author{Hadyan Luthfan Prihadi}

\email{hady001@brin.go.id}
\affiliation{Research Center for Quantum Physics, National Research and Innovation Agency (BRIN), South Tangerang 15314, Indonesia.}
\author{Rafi Rizqy Firdaus}
\email{firdaus.rafi.rizqy.n7@s.mail.nagoya-u.ac.jp}
\affiliation{Department of Physics, Nagoya University,\\
Furo-cho, Chikusa-ku, Nagoya 464-8602, Japan}
\author{Fitria Khairunnisa}
\email{30223301@mahasiswa.itb.ac.id}
\affiliation{Theoretical High Energy Physics Group, Department of Physics, FMIPA, Institut Teknologi Bandung, Jl. Ganesha 10 Bandung, Indonesia.}
\affiliation{Research Center for Quantum Physics, National Research and Innovation Agency (BRIN), South Tangerang 15314, Indonesia.}
\author{Donny Dwiputra}
\email{donny.dwiputra@apctp.org}
\affiliation{Asia Pacific Center for Theoretical Physics, Pohang University of Science and Technology, Pohang 37673, Gyeongsangbuk-do, South Korea.}
\affiliation{Research Center for Quantum Physics, National Research and Innovation Agency (BRIN), South Tangerang 15314, Indonesia.}
\author{Freddy Permana Zen}
\email{fpzen@itb.ac.id}
\affiliation{Theoretical High Energy Physics Group, Department of Physics, FMIPA, Institut Teknologi Bandung, Jl. Ganesha 10 Bandung, Indonesia.}
\affiliation{Indonesia Center for Theoretical and Mathematical Physics (ICTMP), Institut Teknologi Bandung, Jl. Ganesha 10 Bandung,
	40132, Indonesia.}
\date{\today}
\begin{abstract}
    We consider a stationary solution of a charged black hole with scalar hair in AdS$_4$, where the scalar field is coupled to a $U(1)$ Maxwell gauge field. Near the singularity, the spacetime transitions into a more general Kasner geometry. The black hole is then injected with rotating and charged gravitational shock waves in the Dray-'t Hooft solution. These shock waves lengthen the wormhole connecting the two asymptotic boundaries, thereby disrupting the correlations between them. The correlation, quantified by the quantum mutual information between subregions on the left and right boundaries, vanishes at a characteristic timescale known as the scrambling time, which depends logarithmically on the black hole entropy. The mutual information is computed holographically using the Ryu-Takayanagi prescription for entanglement entropy. We investigate how the rotation and charge of both the black hole and the shock waves affect chaotic properties such as the scrambling time delay and the Lyapunov exponent. The interaction between the charges of the black hole and the shock waves introduces a delay in the scrambling process. We find that as the strength of the boundary deformation increases, both the Lyapunov exponent and the scrambling time delay decrease monotonically. Furthermore, the angular momentum of the shock waves enhances both the Lyapunov exponent and the scrambling time delay.
\end{abstract}
\maketitle

\tableofcontents

\section{Introduction}
The study on the interior structure of black hole solutions with scalar hair has continued to attract significant attention until very recently \cite{Frenkel2020,Hartnoll2021,Sword2022a,Sword2022b,Auzzi2022,Cai2023igv,Cai_2021,Are_n_2024,Caceres2024,Caceres:2024edr,Gao2024,Oling2025,Carballo2025,Zhang2025,zhang2025interiorstructureholographics,prihadi2025scramblingchargedhairyblack,Cai2025Kasner}. Previously, the interior structure of a stationary, instead of static, hairy black hole solution in AdS$_3$ has been investigated in \cite{Gao2024}. In this work, we extend the black hole solution to the stationary charged hairy black hole in 4-dimensional AdS by coupling the scalar field in the bulk with a $U(1)$ gauge field $A_\mu$. The charged scalar field generates a holographic renormalization group flow to a more general Kasner spacetime inside the horizon. On the other hand, in the context of the AdS/CFT correspondence \cite{Maldacena1999,witten1998antisitterspaceholography,witten2002multitraceoperatorsboundaryconditions,de_Haro_2001,Bianchi2002}, the scalar field plays a role as the source of a deformation in the boundary CFT theory. This deformation controls the value of the Kasner exponent which characterizes the geometry inside the black hole horizon.\\
\indent The stationary black hole solution has a Killing vector $\zeta_x=\partial_x$ associated with a conserved momentum in the $x$ direction, similar to the angular momentum conservation of a rotating black hole in the spherically symmetric background. One can then add rotating gravitational shock waves with angular momentum $\mathcal{L}$ to the black hole metric yielding the Dray-'t Hooft solution \cite{DRAY1985173,Dray1985}. Rotating shock waves enhance chaos in rotating black holes, as previously studied in the rotating BTZ black hole \cite{Malvimat2022}, Kerr-AdS black hole \cite{Malvimat2023,Malvimat2023b}, and the Kerr-Sen-AdS black hole \cite{Prihadi2023,Prihadi2024}. In this work, we study how the angular momentum of the gravitational shock waves affects the chaotic properties of the stationary charged hairy black hole in AdS$_4$. The chaotic aspects of various black holes have been studied intensively in \cite{Leichenauer_2014,Jahnke_2018,Jahnke2019chaosbound,Malvimat2022,Malvimat2023,Malvimat2023b,Prihadi2023,Prihadi2024,Lilani2025}including static black holes with scalar hair \cite{Caceres2024,prihadi2025scramblingchargedhairyblack}. However, to the best of our knowledge, studies on the chaotic properties of stationary black holes with charged scalar hair are still lacking.\\
\indent The chaotic properties of the black hole can be studied by investigating the exponential decay of the out-of-time order correlator (OTOC) \cite{Shenker2014,Shenker:2014cwa,jahnke2019,Trunin2021}. In studying the decay of the OTOC, we calculate the quantum mutual information $I(A:B)=S_A+S_B-S_{A\cup B}$, where $S_A$ is the entanglement entropy of a subsystem $A$. In this case, $A$ and $B$ are subsystems in the left and right asymptotic boundaries, respectively. The mutual information is calculated using the Ryu-Takayanagi holographic entanglement entropy formula \cite{Ryu2006PRL,Ryu2006JHEP}. The minimal surface corresponding to $S_{A\cup B}$ stretches from the left to the right asymptotic boundary, similar to the Hartman-Maldacena surface \cite{Hartman2013}. This surface is sensitive to gravitational shock waves and slightly penetrates the black hole interior. One can show that the mutual information depends on the time at which the gravitational shock waves are sent from the asymptotic boundary until they reach the horizon at $t\approx 0$. The correlation vanishes at the time scale when the insertion time is $t_*\sim\frac{1}{\lambda_L}\log S$, where $S$ is the black hole's entropy and $\lambda_L$ is the quantum Lyapunov exponent. This concludes that the black hole is a fast scrambler \cite{Sekino_2008} and manifests chaotic behavior.\\
\indent Even though we study the stationary solution to a charged hairy black hole to investigate the chaotic properties of the black hole, this solution might also be important in the context of holographic superconductors \cite{Hartnoll2008,Hartnoll2008PRL}. A phase transition usually occurs in a hairy black hole background with a scalar field $\phi$ coupled with a Maxwell $U(1)$ gauge field. Recently, numerous works \cite{Hartnoll2021,Sword2022a,Sword2022b,Zhang2025,zhang2025interiorstructureholographics} have investigated the interior structure of holographic superconductor models, which posit several interesting phenomena, including the Einstein-Rosen bridge collapse, Josephson oscillation, and Kasner inversion. Our work can serve as a starting point for investigating the role of the rotation parameter in the stationary black hole solution, which relates to the superconductive behavior of the dual field theory.\\
\indent The structure of this paper is as follows. In Section II, we numerically obtain the stationary solution to the charged hairy black hole in AdS$_4$ using AdS boundary conditions. In the bulk, we consider a scalar field coupled to the Maxwell field. We demonstrate that an inner horizon does not form as the metric undergoes a collapse of the Einstein-Rosen bridge. Additionally, we analyze the solution near the singularity, where the geometry approaches a more general Kasner spacetime. We study how the boundary deformation parameter controls the Kasner exponents and examine their dependence on the rotation parameter of the black hole. In Section III, we study the scrambling of information in this black hole background by injecting rotating and charged shock waves. We present holographic calculations of the mutual information and explore the relationship between chaotic parameters (Lyapunov exponents and scrambling time delay) and the boundary deformation, as well as the Kasner exponents. In Section IV, we provide a summary and discussion of the results.
\section{Stationary Charged Hairy Black Hole Solution in AdS$_4$}
In this section, we find a four-dimensional stationary asymptotically AdS charged black hole solution with scalar field $\phi$ coupled with Maxwell field $A_\mu$. The action with both AdS radius $L$ and the gravitational coupling $\frac{1}{16\pi G_N}$ are set to one is given by
\begin{align}
    S=\int d^4x\sqrt{|g|}\bigg(&R-2\Lambda-g^{\mu\nu}(\partial_\mu-iqA_\mu)\phi(\partial_\nu+iqA_\nu)\phi^*-m^2|\phi|^2-\frac{1}{4}F_{\mu\nu}F^{\mu\nu}\bigg),
\end{align}
where $F_{\mu\nu}=\partial_\mu A_\nu-\partial_\nu A_\mu$ is the field strength tensor of a $U(1)$ gauge field $A_\mu$ and $\phi$ is in general a complex scalar field. Here, we have set the AdS radius and the gravitational coupling equal to one. The cosmological constant is negative and it is written as $\Lambda=-3$. One of the equations of motion for the Maxwell field states that the complex phase of the scalar field $\phi$ is a constant. Therefore, we may choose a gauge so that the phase is zero and the scalar field becomes real without changing the equations of motion. In this work, we consider a radially-dependent scalar field
\begin{equation}
    \phi=\phi^*=\phi(r).
\end{equation}
\indent The stationary metric ansatz where the AdS radius $L$ is set to $L=1$ and following the symmetry $t\rightarrow-t$ and $x\rightarrow-x$ is given by
\begin{align}
    ds^2=\frac{1}{r^2}\qty(-f(r)e^{-\chi(r)}dt^2+\frac{dr^2}{f(r)}+(N(r)dt+dx)^2+dy^2),
\end{align}
where $r$ denotes the AdS radial coordinate with $r\rightarrow0$ is the asymptotic AdS boundary and $r=r_H$ is the horizon radius satisfying $f(r_H)=0$. Throughout the numerical calculations, we choose $r_H=1$. For a stationary solution, we also choose the non-zero components of the Maxwell gauge field $A_\mu$ to be
\begin{equation}
    A_\mu dx^\mu=A_t(r)dt+A_x(r)dx.
\end{equation}
The addition of the $x$ component, in contrast with the previous non-rotating charged hairy black hole solution such as in \cite{prihadi2025scramblingchargedhairyblack}, is crucial in generating the stationary solution with $N(r)\neq0$. A similar stationary ansatz with nonvanishing $A_t$ and $A_x$ has also been employed in \cite{bravogaete2025rotatingaxionicads4black} to study axionic black holes in AdS$_4$.\\
\indent All of the fields $\{\phi,A_t,A_x,f,\chi,N\},$ are obtained numerically as the solution to the equations of motion. In this case, the equations of motion are the Einstein's equations
\begin{equation}
    R_{\mu\nu}-\frac{1}{2}Rg_{\mu\nu}-\Lambda g_{\mu\nu}=T_{\mu\nu},
\end{equation}
with
\begin{align}
    T_{\mu\nu}=&\frac{1}{2}F_{\mu\alpha}F_\nu^{\alpha}-\frac{1}{8}g_{\mu\nu}F_{\alpha\beta}F^{\alpha\beta}-\frac{g_{\mu\nu}}{2}(\nabla_\alpha-iqA_\alpha)\phi(\nabla^\beta+iqA^\beta)\phi^*\\\nonumber
    &-\frac{g_{\mu\nu}}{2}m^2|\phi|^2+\frac{1}{2}((\nabla_\mu-iqA_\mu)\phi(\nabla_\nu+iqA_\nu)\phi^*+\mu\leftrightarrow\nu),
\end{align}
the Maxwell's equations
\begin{equation}
    \nabla^{\mu} F_{\mu\nu}=iq(\phi^*(\nabla_\nu-iqA_\nu)\phi-\phi(\nabla_\nu+iqA_\nu)\phi^*),
\end{equation}
and the Klein-Gordon equation
\begin{equation}
    (\nabla_\mu-iqA_\mu)(\nabla^\mu-iqA^\mu)\phi-m^2\phi=0,
\end{equation}
where $\nabla_\mu$ denotes the covariant derivative with respect to the metric $g_{\mu\nu}$. In this work, we use $m^2=-2$, which satisfies the Breitenlohner-Freedman bound. \\
\indent The equations of motion give us the following coupled differential equations. The $t-$component of the Maxwell's equations gives us
\begin{align}\label{EoMAt}
    &{A_t}''+{A_t}' \left(\frac{{N} e^{\chi} {N}'}{f}+\frac{\chi '}{2}\right)-\frac{2 q^2 {A_t} \phi^2}{r^2 f}=A_x'N\bigg(\frac{-f'}{f}+\chi'+\frac{Ne^\chi N'}{f}\bigg)+A_x'N',
\end{align}
while the $x-$component gives us
\begin{align}\label{EoMAx}
    &{A_x}''+A_x'\bigg(\frac{f'}{f}-\frac{\chi'}{2}-\frac{Ne^\chi N'}{f}\bigg)-\frac{2 q^2 {A_x} \phi^2}{r^2 f}=-\frac{e^{\chi} {A_t}' {N}'}{f}.
\end{align}
The Klein-Gordon equation is given by
\begin{align}\label{EoMphi}
    \phi''+\phi' &\left(\frac{f'}{f}-\frac{\chi'}{2}-\frac{2}{r}\right)+\phi\bigg(-\frac{2 q^2 A_tA_xNe^{\chi}}{f^2}+\frac{q^2A_t^2 e^{\chi}}{f^2}+\frac{q^2A_x^2N^2 e^{\chi}}{f^2}-\frac{q^2A_x^2}{f}-\frac{m^2}{r^2 f}\bigg)=0.
\end{align}
The $tx-$component of the Einstein's equations gives us the coupled second-order differential equation for $N(r)$ as follows
\begin{align}\label{EoMEins1}
    {N}''+&{N}' \left(\frac{\chi'}{2}-\frac{2}{r}\right)+{N} \left(-r^2{A_x}'^2-\frac{2 q^2{A_x}^2 \phi^2}{f}\right)+r^2 {A_t}'{A_x}'+\frac{2 q^2 {A_t} {A_x} \phi^2}{f}=0.
\end{align}
Lastly, the final two equations are the remaining Einstein's equations that give us
\begin{align}\label{EoMEins2}
    &-\frac{r^2{N} e^{\chi} {A_t}'{A_x}'}{f}+\frac{r^2 e^{\chi}{A_t}'^2}{2 f}-\frac{2 q^2{A_t}{A_x}{N} e^{\chi} \phi^2}{f^2}+\frac{q^2{A_t}^2 e^{\chi} \phi^2}{f^2}\\\nonumber
    &+\frac{r^2{N}^2 e^{\chi} {A_x}'^2}{2 f}+\frac{1}{2} r^2{A_x}'^2+\frac{q^2{A_x}^2 {N}^2 e^{\chi} \phi^2}{f^2}+\frac{q^2 {A_x}^2 \phi^2}{f}\\\nonumber
    &-\frac{2 f'}{r f}+\frac{m^2 \phi^2}{r^2 f}+\frac{e^{\chi} {N}'^2}{2 f}-\frac{6}{r^2 f}+\frac{6}{r^2}+\phi'^2=0, 
\end{align}
and
\begin{align}\label{EoMEins3}
    -&\frac{2 q^2{A_t}{A_x}{N}e^{\chi} \phi^2}{f^2}+\frac{q^2{A_t}^2 e^{\chi} \phi^2}{f^2}+\frac{1}{2} r^2 {A_x}'^2+\frac{q^2{A_x}^2{N}^2 e^{\chi} \phi^2}{f^2}-\frac{\chi '}{r}+\phi'^2=0. 
\end{align}
These equations of motion can be simplified as
\begin{align}
    e^{-\chi/2}(e^{\chi/2}A_t')'&=\frac{2q^2A_t\phi^2}{r^2f}-\frac{A_x'N^2e^{\chi}}{f}\bigg(\frac{fe^{-\chi}}{N}\bigg)'+\frac{e^\chi(N^2)'}{2}(A_x'N-A_t'),\\
    e^{\chi/2}(fe^{-\chi/2}A_x')'&=\frac{2q^2A_x\phi^2}{r^2}-e^\chi N'(A_t'-A_x'N),\\
    r^2e^{\chi/2}\bigg(\frac{fe^{-\chi/2}\phi'}{r^2}\bigg)'&=\phi\bigg(\frac{m^2}{r^2}+q^2A_x^2-\frac{q^2e^\chi}{f}(A_t-A_xN)^2\bigg),\\
    r^2e^{-\chi/2}\bigg(\frac{e^{\chi/2}N'}{r^2}\bigg)'&=r^2A_x'(A_x'N-A_t')+\frac{2q^2\phi^2A_x}{f}(A_xN-A_t),\\
    \frac{\chi'}{r}&=\frac{q^2\phi^2e^\chi}{f^2}(A_t-A_xN)^2+\frac{r^2}{2}A_x'^2+\phi'^2,\\
    4r^4e^{\chi/2}\bigg(\frac{fe^{-\chi/2}}{r^3}\bigg)'&=2m^2\phi^2-12+2r^2q^2A_x^2\phi^2+r^2e^\chi(N'^2+r^2(A_t'-A_x'N)^2).
\end{align}
\indent We solve these equations of motion numerically, subject to the AdS boundary conditions. When both $N$ and $A_x$ are zero, the equations reduce to the set of equations of motion for a charged hairy black hole in AdS$_4$, which is commonly used to study holographic superconductors \cite{Hartnoll2008,Hartnoll2021}. Note that one cannot simply recover the rotating hairy black hole equations of motion, such as in \cite{Gao2024} when the gauge field $A_\mu$ vanishes. This is because here, we use a real scalar field while the aforementioned work uses a complex scalar field. We also see that, when $A_x=0$, Eq. \eqref{EoMAx} forces us to have a constant value of $N(r)$. By setting the boundary condition $N(0)=0$, we have $N(r)=0$ everywhere. Therefore, stationary solution with $N(r)\neq0$ requires $A_x(r)\neq0$. 
\subsection{Near-boundary and near-horizon expansions}
\begin{figure}
    \centering
    \includegraphics[scale=0.37]{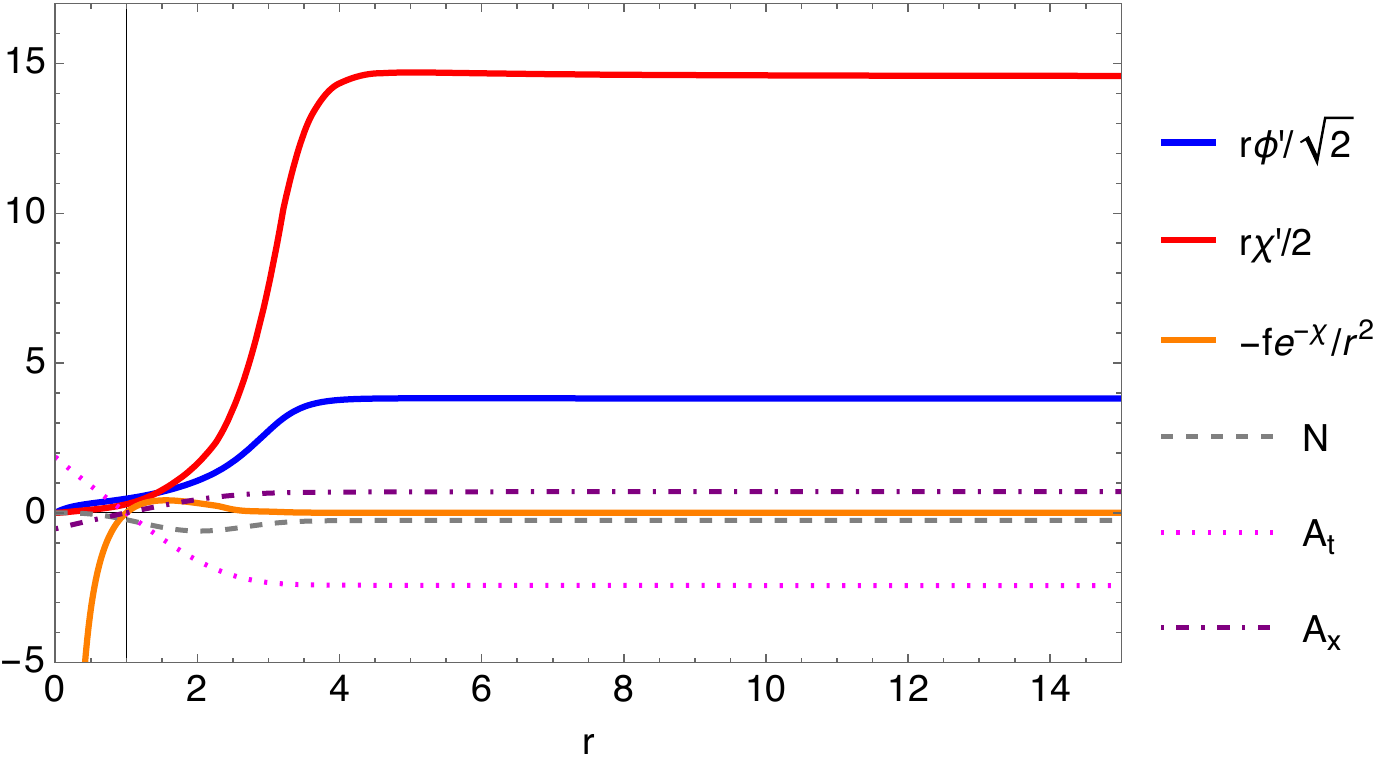}
    \includegraphics[width=0.44\linewidth]{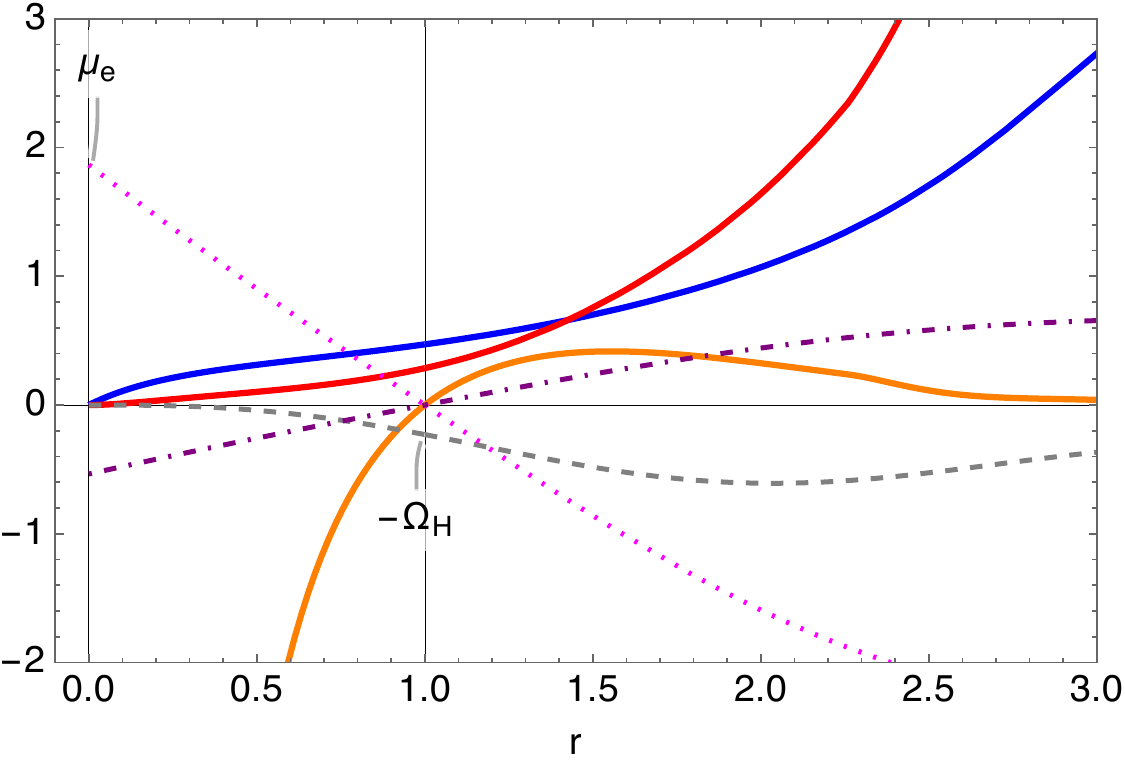}
    \caption{Numerical solutions at $\phi_0/T=6.48714$. In this plot, we use $q=0.10, N_{H1}=-A_{xH1}=-1/2, \mathcal{L}=0.1,$ and $\rho=2$. The horizon is located at $r_H=1$, while the boundary solutions are represented by $r\rightarrow0$.}
    \label{fig:solutions}
\end{figure}

In solving the equations of motion numerically, we need to specify boundary conditions. The differential equations are second order in $\phi,A_t,A_x,N$ and first order in $f,\chi$. We integrate the equations of motion numerically by dividing them into two parts. The first part is the integration from the near-horizon region with $r=r_H-\delta$ to the boundary at $r\rightarrow 0$ while the second part starts from $r=r_H+\delta$ to the singularity where $r$ is large and $\delta$ is a small number. In this work, we choose $\delta=10^{-5}$ throughout the numerical calculations. Furthermore, we also choose $m^2=-2$, which satisfies the Breitenlohner–Freedman bound and generates relevant deformation in the CFT.\\
\indent We need to specify the values of the fields near the horizon as the initial conditions to the numerical calculations. The expansions of the fields near the horizon are given by
\begin{align}
    \phi(r)=&\phi_{H0}+\phi_{H1}(r-r_H)+...\;,\label{nearhorizonphi}\\
    f(r)=&f_{H1}(r-r_H)+...\;,\label{nearhorizonf}\\
    A_{t}(r)=&A_{tH1}(r-r_H)+A_{tH2}(r-r_H)^2+...\;,\label{nearhorizonat}\\
    A_x(r)=&A_{xH1}(r-r_H)+A_{xH2}(r-r_H)^2+...\;,\label{nearhorizonax}\\
    \chi(r)=&\chi_{H0}+\chi_{H1}(r-r_H)+...\;,\label{nearhorizonchi}\\
    N(r)=&N_{H0}+N_{H1}(r-r_H)+...\;.\label{nearhorizonn}
\end{align}
The value of $f(r)$ at $r_H$ is set to zero for the black hole solution, while $f_{H1}$ is related to the black hole's temperature through the relation
\begin{equation}
    T=\frac{|f'(r_H)|e^{-\chi(r_H)/2}}{4\pi}.
\end{equation}
Furthermore, to make the norm of the gauge field $A_\mu$ finite at the horizon, we also require $A_t(r_H)=A_x(r_H)=0$. The value of $N(r)$ evaluated at the horizon is nothing but the horizon's angular velocity $N(r_H)=N_{H0}=-\Omega_H$, with
\begin{align}
    N_{H0}=&\frac{A_{tH1}A_{xH1}+2N_{H1}}{A_{xH1}^2}\\\nonumber
    &+\frac{e^{-\text{$\chi_{H0}$}} \sqrt{-A_{xH1}^4N_{H1}^2 e^{2 \text{$\chi_{H0}$}}+12A_{xH1}^4 e^{\text{$\chi_{H0}$}}+4A_{xH1}^4 e^{\text{$\chi_{H0}$}} \text{$\phi_{H0}$}^2+4A_{xH1}^2N_{H1}^2 e^{2 \text{$\chi_{H0}$}}}}{A_{xH1}^3}.
\end{align}
We then substitute these expansions to the equations of motion to find the expansion coefficients by setting each order of expansion to zero. It turns out that only 5 parameters are free. We choose $\phi_{H0},\chi_{H0},A_{tH1},A_{xH1},N_{H1}$ as the free parameters, subject to the desired AdS boundary condition at $r\rightarrow 0$. This can be achieved using the shooting method as in \cite{Hartnoll2008,Hartnoll2021}. The explicit form of the other expansion coefficients are written in the Appendix A.\\
\indent The expansions of the fields near $r\rightarrow0$, where the CFT lives, are given by
\begin{align}
    \phi(r)=&\phi_{b1}r+\phi_{b2}r^2+\frac{1}{4} \text{$\phi_{b1}$} \left(2 q^2 \left(A_{xb0}^2-A_{tb0}^2 e^{\text{$\chi_{b0}$}}\right)+\text{$\phi_{b1}$}^2\right)r^3+...\;,\\
    f(r)=&1+\frac{1}{2}\phi_{b1}^2r^2+...\;,\\
    A_t(r)=&A_{tb0}+A_{tb1}r+A_{tb0}q^2\phi_{b1}^2r^2+\frac{1}{12} \text{$\phi_{b1}$} \left(8A_{tb0} q^2 \text{$\phi_{b2}$}+A_{tb1} \left(4 q^2-1\right) \text{$\phi_{b1}$}\right)r^3+...\;,\\
    A_x(r)=&A_{xb0}+A_{xb1}r+A_{xb0}q^2\phi_{b1}^2r^2+\frac{1}{12} \text{$\phi_{b1}$} (8A_{xb0} q^2 \text{$\phi_{b1}$}+A_{xb1}(4 q^2-1) \text{$\phi_{b1}$})r^3+...\;,\\
    \chi(r)=&\chi_{b0}+\frac{\phi_{b1}^2}{2}r^2+\frac{4\phi_{b1}\phi_{b2}}{3}r^3+...\;,\\
    N(r)=&N_{b0}+\frac{1}{4} \left(2A_{xb0}q^2 \text{$\phi_{b1}$}^2 (A_{xb0}N_{b0}-A_{tb0})-A_{tb1}A_{xb1}+A_{xb1}^2N_{b0}\right)r^4+...\;.
\end{align}
This expansion is derived for the specific choice $m^2=-2$, while the generalization to arbitrary $m^2$ is straightforward.\\
\indent The coefficients of the boundary expansions are related to the CFT data. For example, in the AdS$_4$/CFT$_3$ dictionary with $m^2=-2$, it is known that $\phi_{b1}$ plays a role as the source to the boundary deformation denoted as $\phi_0$, which deform the theory with an action in the form of
\begin{equation}
    \delta S_{CFT}=\int d^3x\phi_0\mathcal{O},
\end{equation}
where $\mathcal{O}$ is the response to the deformation. From the holographic renormalization procedure \cite{Bianchi_2002}, the single-trace operator $\mathcal{O}$, or the response to the deformation, is given by $\phi_{b2}$. Furthermore, $A_{tb0}=\mu_e$ and $A_{xb0}=\mu_m$ represent the boundary chemical potential associated with the electric and magnetic field, respectively, while $A_{tb1}=\rho_e$ and $A_{xb1}=J_x$ represent the electric charge density and current density operator, respectively. See, for example, \cite{bravogaete2025rotatingaxionicads4black} for the usage of $A_x$ component in AdS/CFT.\\
\indent In an asymptotically AdS spacetime, $\chi$ needs to vanish at $r\rightarrow 0$. By choosing an arbitrary value of $\chi$ at the horizon as the initial condition, $\chi(0)=0$ does not automatically satisfied. However, using the symmetry
\begin{equation}
    e^\chi\rightarrow a^2e^\chi,\;\;\;t\rightarrow at,\;\;\;A_t\rightarrow\frac{A_t}{a},\;\;\;N\rightarrow\frac{N}{a},\label{symmetry1}
\end{equation}
we can rescale the fields by choosing $a$ so that $\chi(0)=0$ is satisfied. Therefore, we can set $\chi_{H0}=1$ throughout the calculations. Furthermore, we also want the rotation to vanish at the boundary, i.e. $N(0)=0$. Similar to $\chi(r)$, this is also not automatically satisfied for an arbitrary value of $N_{H1}$. However, the equations of motion are also invariant under the shift
\begin{equation}
    N\rightarrow N+b,\;\;\;x\rightarrow x-bt,\;\;\;A_t\rightarrow A_t+bA_x.\label{symmetry2}
\end{equation}
Therefore, we can choose the value of $b$ such that $N(0)=0$ is satisfied. \\
\indent We then solve the equations of motion numerically using the shooting method. First, we fix the charge density $\rho$, which determines the black hole's charge. Next, we choose a value for $\phi_{H0}$ and set $\chi_{H0} = 1$. This choice determines the value of $\phi_0$ at the boundary. We also specify the angular velocity of the black hole by fixing $N_{H1}$ and setting $A_{xH1} = -N_{H1}$. Subsequently, we apply the transformations given in Eqs.~\eqref{symmetry1} and \eqref{symmetry2} to shift all fields such that $\chi(0) = 0$ and $N(0) = 0$. Finally, we determine the value of $A_{tH0}$ using the shooting method so that the boundary condition $A_t'(0) = -\rho$ is satisfied. The numerical solutions to the equations of motion are shown in Figure \ref{fig:solutions}. These solutions give us $\mu_e$ from $A_t(0)$ and $-\Omega_H$ from $N(r_H)$. Furthermore, the solutions to $\phi(r)$ and $\chi(r)$ also give us the constant $c$ at the interior for determining the Kasner exponents, while the solutions for $N(r),A_t(r),A_x(r)$ determine their constants near the singularity. 
\subsection{Einstein-Rosen Bridge Collapse and Ergosphere}
In this section, we numerically show that the function $-\frac{fe^{-\chi}}{r^2}$ exponentially contracts right after the would-be inner horizon. The fact that $f(r)$ does not become exactly zero at $r>r_H$ tells us that the inner horizon does not form. Furthermore, we also see how the scalar field affects the ergosphere, in which its radius is defined when the function
\begin{equation}
    g_{tt}(r)=-\frac{fe^{-\chi}+N^2}{r^2},
\end{equation}
vanishes.
\subsubsection{Einstein-Rosen Bridge Collapse}
\begin{figure}
    \centering
    \includegraphics[width=0.6\linewidth]{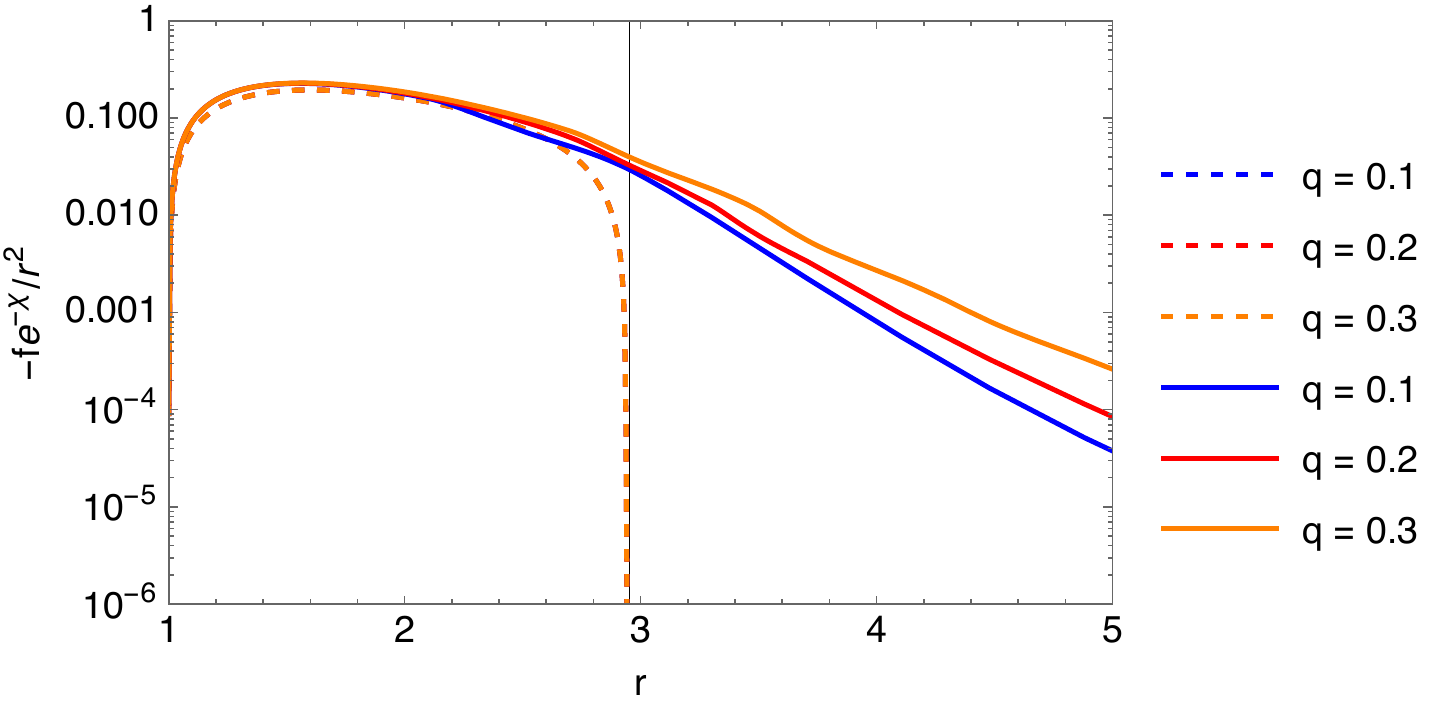}
    \caption{Collapse of the Einstein-Rosen bridge when the coupling $q$ is varied. Here, we use $N_{H1}=-1/2, \rho=2$, and $\phi_0/T=5.59021$. The ``would-be" inner horizon when $\phi=0$ is located at $r\approx2.95$.}
    \label{fig:ercollapseq}
\end{figure}

The function $-\frac{fe^{-\chi}}{r^2}$ becomes zero exactly at the outer horizon $r_H$. When $\phi=0$, we expect that a stationary charged black hole have another horizon located at $r= r_I>r_H$, when $f(r_I)=0$, which is called the inner horizon. Therefore, we also expect that the function $-\frac{fe^{-\chi}}{r^2}$ becomes precisely zero at the inner horizon as well. However, it turns out that black holes with $\phi\neq0$ does not have an inner horizon even if it is charged or rotating. Instead, the function $-\frac{fe^{-\chi}}{r^2}$ exponentially decays to zero after $r=r_I$, which is the ``would-be" inner horizon when $\phi=0$. This decay is called the Einstein-Rosen bridge collapse, first studied in \cite{Hartnoll_2020}.\\
\begin{figure}
    \centering
    \includegraphics[width=0.6\linewidth]{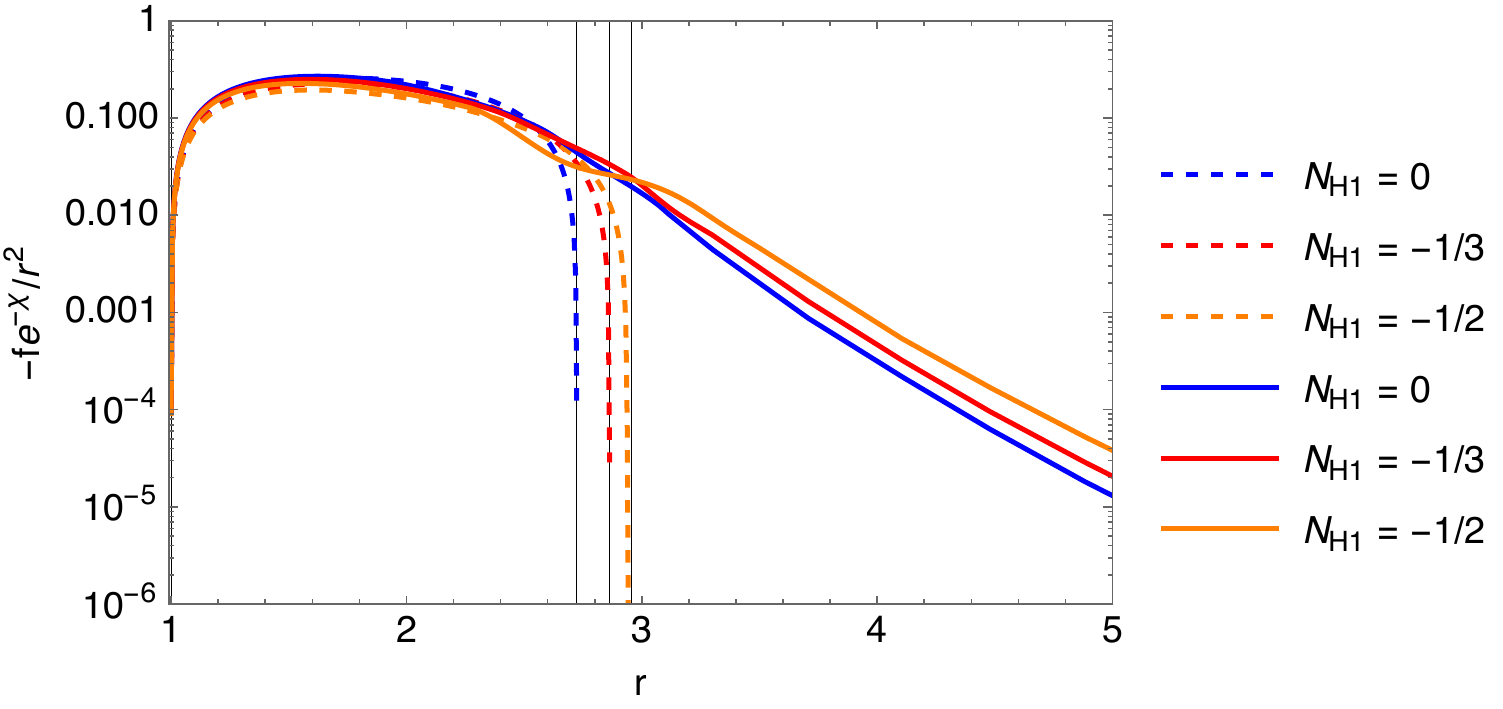}
    \caption{Collapse of the Einstein-Rosen bridge when the rotation parameter $N_{H1}$ is varied. Here, we use $q=0.10,\rho=2$, and $\phi_0/T=6.48714$. The dashed lines represent the solutions when $\phi=0$. The ``would-be" inner horizons are located at $r\approx2.72$ for $N_{H1}=0$, $r\approx2.86$ for $N_{H1}=-1/3$, and $r\approx2.95$ for $N_{H1}=-1/2$.}
    \label{fig:ercollapse}
\end{figure}
\indent The exponential decay of $-\frac{fe^{-\chi}}{r^2}$ after the ``would-be" inner horizon in the stationary charged hairy black hole is shown in Figure \ref{fig:ercollapse}. When the scalar field is absent, the rotation parameter $N_{H1}$ enlarges the inner horizon, bringing it closer to the singularity. The numerical solution to the black hole without scalar hair is shown in Appendix B. When the scalar field is turned on, the rotation parameter also delays the collapse as it increases. Note that, the choice of negative $N_{H1}$ leads to a positive horizon's angular velocity $\Omega_H$ through the relation $N_{H1}=-\Omega_H$. When $N_{H1}$ becomes more negative, $\Omega_H$ becomes larger and more positive. We also see how the coupling $q$ affect the Einstein-Rosen bridge collapse in Figure \ref{fig:ercollapseq}. It also toned down the decay process when $\phi$ is turned on. In \cite{Sword2022a}, it is shown that when the Einstein-Maxwell-Scalar coupling $\gamma$ is turned on, the function $g_{tt}$ grows first after passing the ``would-be" inner horizon, and then decay hereafter. In this work, we do not find such a solution although we may expect similar behavior happens when we include the EMS term to the action.\\
\indent Here, we only numerically show that the inner horizon does not form through the Einstein-Rosen bridge collapse. The analytical proof of no horizon for rotating and charged black holes is beyond the scope of this work. In Appendix C, we show that the method in \cite{Hartnoll_2020} cannot be used for both charged and rotating hairy black holes. There are, however, several other approaches to argue against the existence of an inner horizon in charged hairy black holes, including calculations of the Noether charge \cite{Cai_2021,Hartnoll2021} and considerations based on energy conditions such as the null energy condition \cite{An2021}. When rotation is included, one may expect modifications influenced by angular momentum, and consequently a general analytical argument excluding inner horizons for stationary or rotating black holes remains elusive. A complete analytical understanding of no inner horizon for this black hole background remains an interesting direction for future work.
\subsubsection{The Metric Component $g_{tt}$ and Ergosphere}
\begin{figure}
    \centering
    \includegraphics[width=0.5\linewidth]{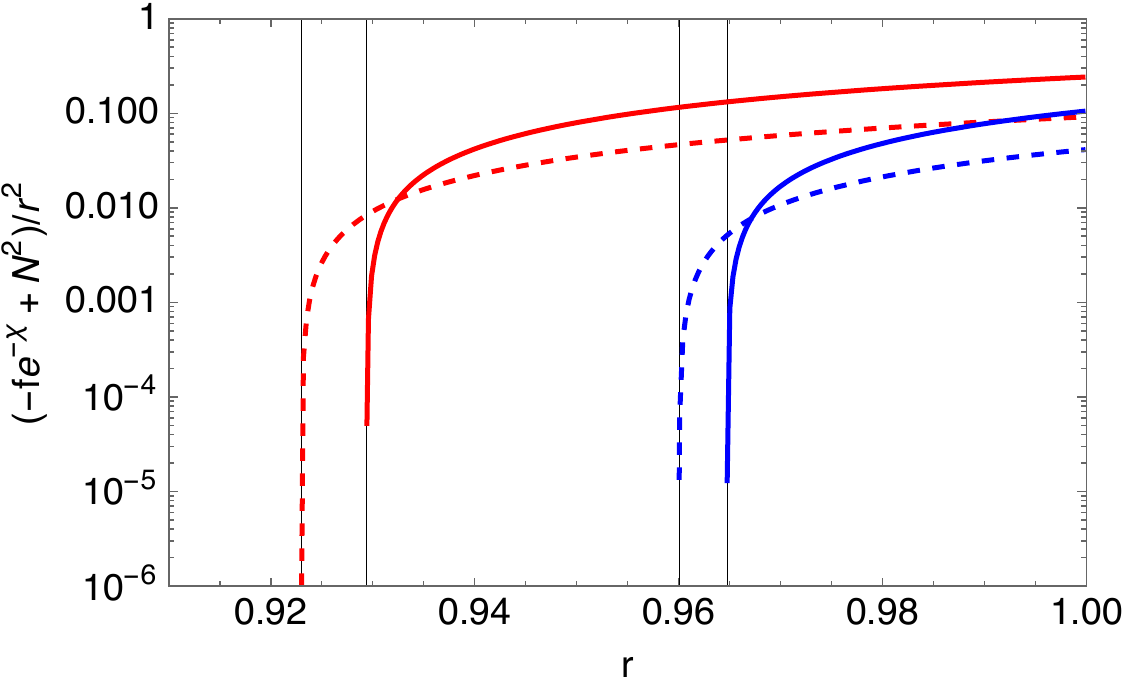}
    \caption{Plot of the metric component $g_{tt}$ as a function of $r$ in the exterior region. The solid lines correspond to the solution with $\phi_0/T = 238.799$ and $q = 0.1$, $\rho = 2$, while the dashed lines represent the case with $\phi = 0$. Blue lines indicate $N_{H1} = -1/2$, and red lines indicate $N_{H1} = -3/4$.}
    \label{fig:ergoouter}
\end{figure}

\begin{figure}
    \centering
    \includegraphics[width=0.5\linewidth]{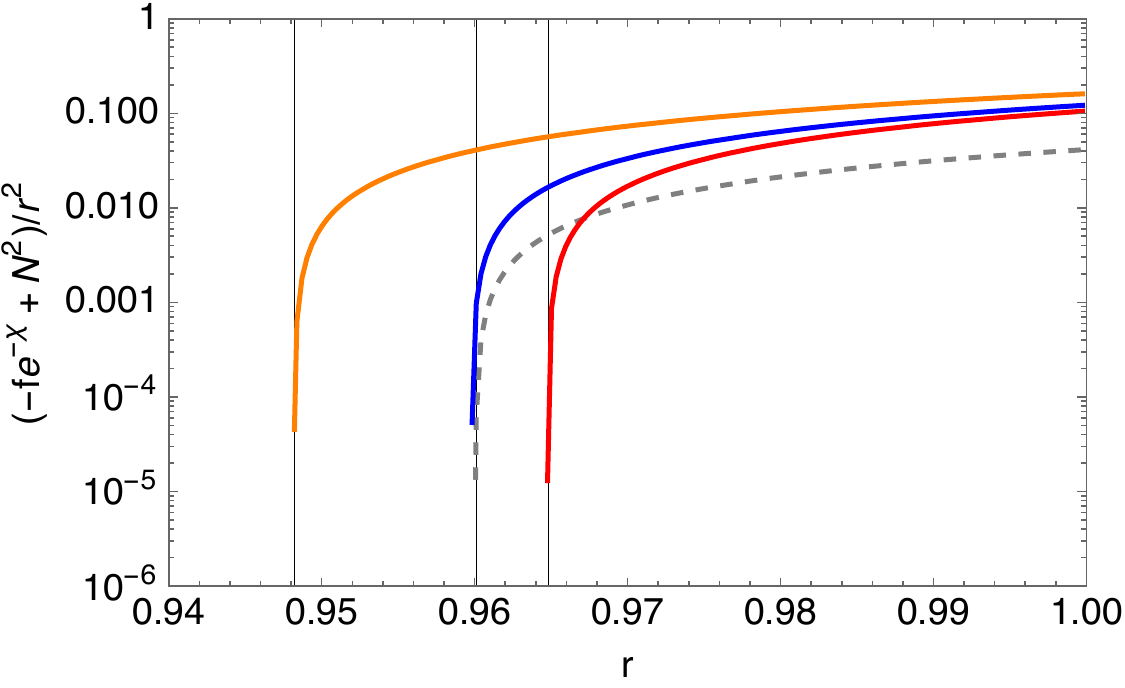}
    \caption{Plot of the metric component $g_{tt}$ as a function of $r$ in the exterior region. The solid lines correspond to the solution with $\phi_0/T=242.795$ and $N_{H1}=-1/2$, $\rho=2$, while the gray dashed line represents the case with $\phi=0$. The red line indicates $q=0.1$, the blue line indicates $q=0.2$, and the orange line indicates $q=0.3$.}
    \label{fig:ergoouterq}
\end{figure}

\begin{figure}
    \centering
    \includegraphics[width=0.5\linewidth]{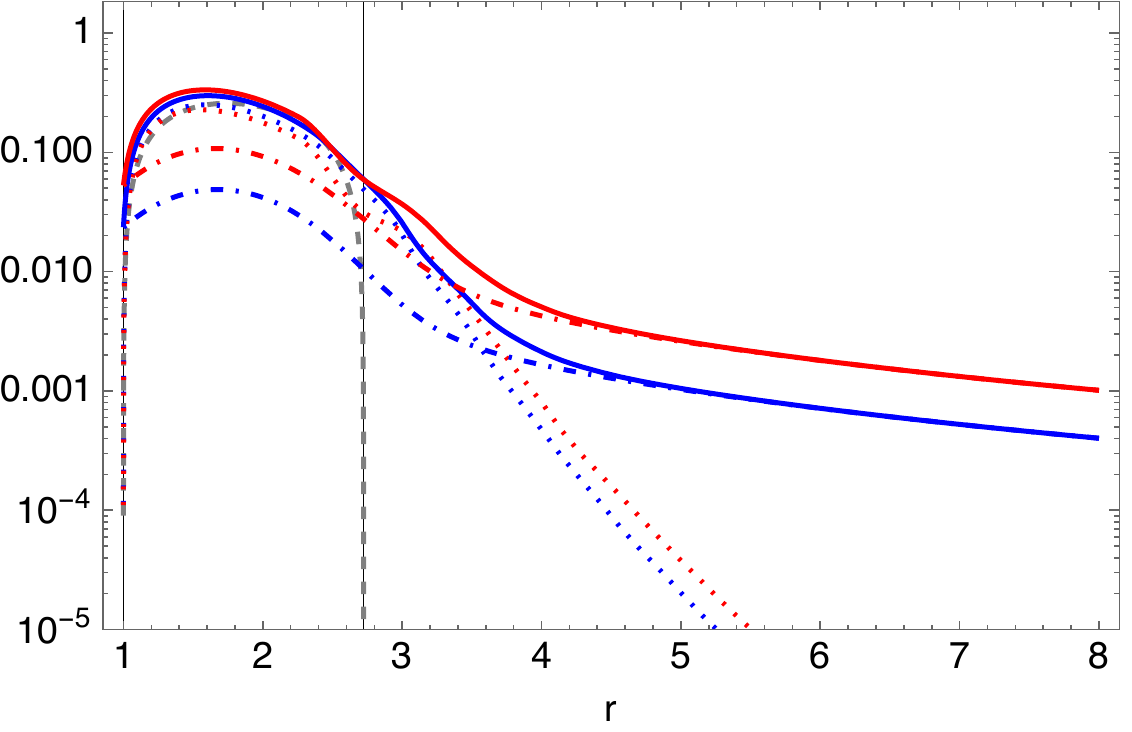}
    \caption{The black dashed line represent the function $-\frac{fe^{-\chi}}{r^2}$ for $\phi=0$, the dotted lines represent the function $-\frac{fe^{-\chi}}{r^2}$ for $\phi\neq0$, the dot-dashed lines represent the function $\frac{N^2}{r^2}$,     and the solid lines are the plots for $g_{tt}(r)$. The blue lines represent $N_{H1}=-1/3$ while the red lines represent $N_{H1}=-1/2$.}
    \label{fig:ergointerior}
\end{figure}

\indent Note that typical rotating black holes have an ergosphere, which is defined as the radius where $g_{tt}=0$. This is because, for rotating black holes, $g_{tt}=0$ does not imply $f=0$, due to the existence of the rotation function $N(r)$. We define the radius with $g_{tt}(r_e)=0$ as the ergosphere's radius $r_e$. The stationary solution to a charged hairy black hole also has an ergosphere in its exterior. One can see in Figure \ref{fig:ergoouter} that the function $g_{tt}$ hits zero not far from the horizon $r_H=1$. In that case, the ergosphere radius for $\phi=0$ case are $r_e\approx0.9601$ for $N_{H1}=-1/2$ and $r_e\approx0.9230$ for $N_{H1}=-3/4$. When the scalar field is turned on, we find that the ergosphere's radius is generally increased, bringing it closer to the horizon. In this case, we have $r_e\approx0.9648$ for $N_{H1}=-1/2$ and $r_e\approx0.9294$ for $N_{H1}=-3/4$.\\
\indent The ergosphere radius exhibits a noticeably different behavior when the coupling constant $q$ is varied for $\phi \neq 0$. The plot of $g_{tt}$ versus $r$ for various values of $q$ is shown in Figure~\ref{fig:ergoouterq}. For $\phi = 0$, the parameter $q$ has no effect on the solution, as it represents the coupling between the scalar field $\phi$ and the Maxwell field $A_\mu$. However, when the scalar field is turned on, the behavior of the ergosphere radius $r_e$ becomes sensitive to the value of $q$. At small $q$, the radius $r_e$ lies closer to the horizon, while for larger values of $q$, it shifts farther away. Specifically, we find $r_e \approx 0.9648$ for $q=0.1$, $r_e \approx 0.9598$ for $q=0.2$ (nearly coinciding with the $\phi=0$ case at $r_e \approx 0.9601$), and $r_e \approx 0.9482$ for $q=0.3$.\\
\indent We may expect to have another ergoregion in the interior of the black hole, i.e., the solution to $g_{tt}=0$ at $r>r_H$, for $\phi=0$. This solution exists for some rotating and charged black holes (see, for example, \cite{Prihadi_2020}). However, in the stationary charged hairy black hole, the solution to $g_{tt}=0$ does not exist in the interior, due to the Einstein-Rosen bridge collapse. We see that, in Figure \ref{fig:solutions}, the function $N(r)$ approaches constant at the deep interior. Therefore, the term $-\frac{fe^{-\chi}}{r^2}$ mainly controls the function $g_{tt}$ in the interior. The plot in Figure \ref{fig:ergointerior} concludes that $g_{tt}$ will not get to zero exactly at some point, rather showing a decay behavior. At large-$r$ (in the Kasner regime), the function $g_{tt}$ decays to zero as $\mathcal{O}(r^{-2})$ with the proportionality constant is given by the constant value of $N(r)$ in the Kasner regime. 
\subsection{Kasner Solution Near the Singularity}
\begin{figure}
    \centering
    \includegraphics[width=0.45\linewidth]{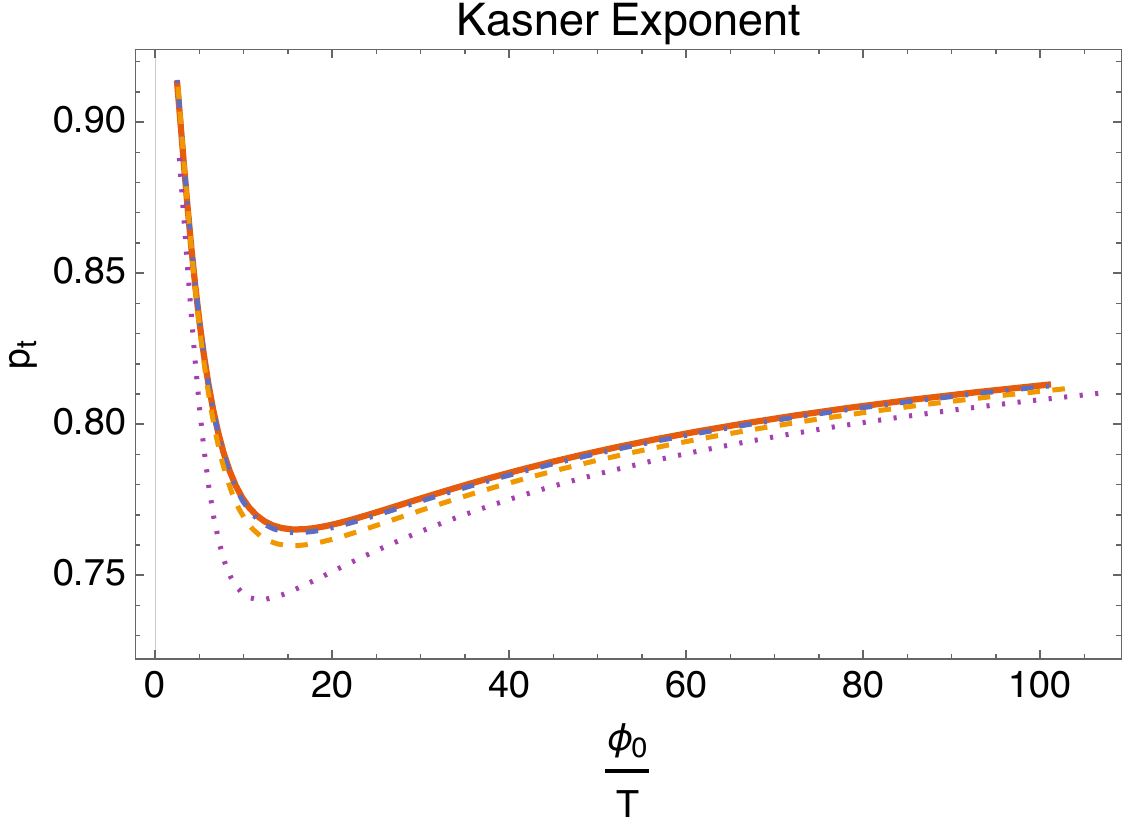}
    \includegraphics[width=0.45\linewidth]{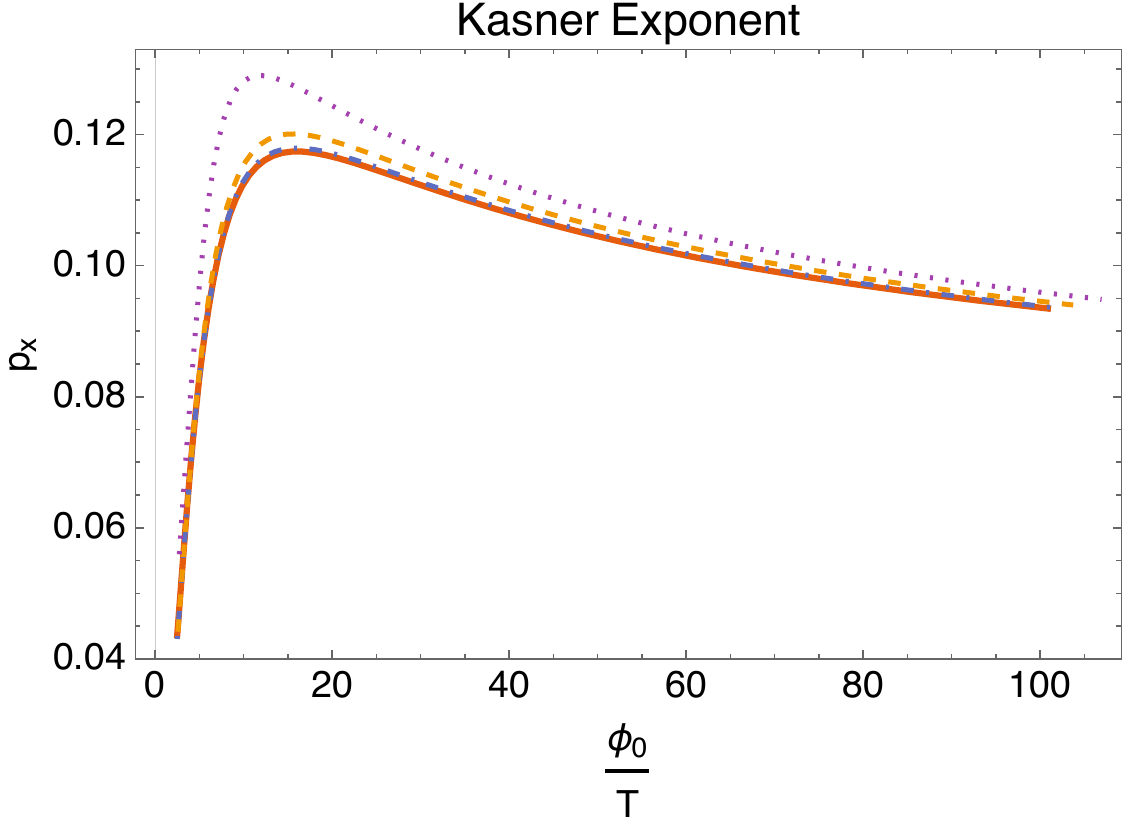}\\
    \includegraphics[width=0.45\linewidth]{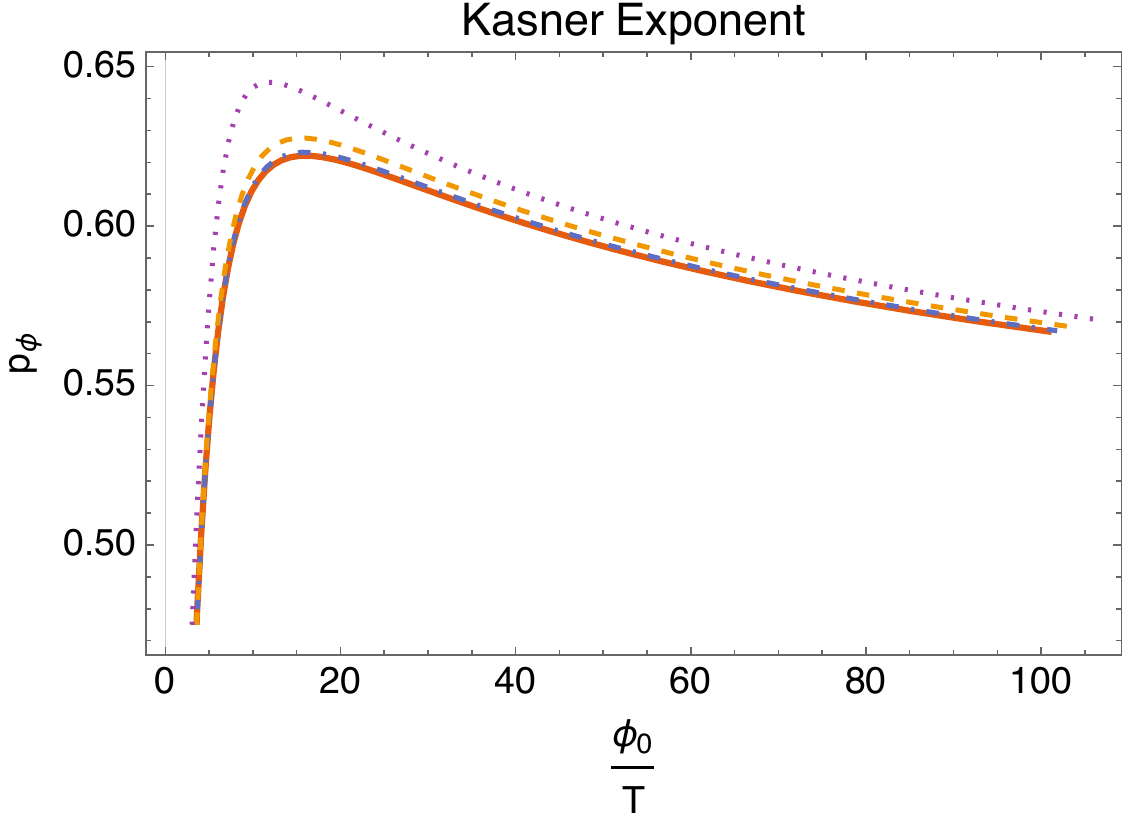}
    \caption{Plot of the Kasner exponents $p_t,p_x,p_\phi$ versus the boundary deformation parameter $\phi_0$ for various $N_{H1}=0$ (solid orange), $N_{H1}=1/4$ (dashed yellow), $N_{H1}=1/2$ (dot-dashed blue), and $N_{H1}=3/4$ (dotted purple). Here, we choose $A_{xH1}=-N_{H1}$ and $q=0.1.$}
    \label{fig:kasnerNH}
\end{figure}

In this section, we explore the solutions to the equations of motion near the singularity. Taking $r\to \infty$ gives us the following equations. First, the Maxwell's equations for $A_t(r)$ becomes
\begin{equation}
\frac{N A_x' f'}{f} + \frac{1}{2} A_t' \chi' - A_x' (N' + N \chi') + A_t'' = 0,\label{eqkasnermaxwt}
\end{equation}
and the Maxwell's equations for $A_x(r)$ becomes
\begin{equation}
A_x''+A_x'\bigg(\frac{f'}{f}-\frac{\chi'}{2}\bigg) = 0.\label{eqkasnermaxwx}
\end{equation}
Furthermore, the Klein-Gordon equation becomes
\begin{equation}
\phi'' + \phi'\left[\frac{f'}{f}- \frac{\chi'}{2} - \frac{2}{r} \right] = 0,\label{eqkasnerscalar}
\end{equation}
while the Einstein's equations become
\begin{equation}
r^2 A_x' (A_t' - N A_x') - \frac{2 N'}{r} + \frac{1}{2} N' \chi' + N'' = 0,\label{eqkasnereinstein1}
\end{equation}
\begin{equation}
\frac{1}{2} r^2 A_x'^2 + \frac{6 }{r^2}+\phi'^2- \frac{2}{r}\frac{ f'}{f} = 0,\label{eqkasnereinstein2}
\end{equation}
and
\begin{equation}
\frac{1}{2}r^2A_x'^2 - \frac{\chi'}{r} + \phi'^2 = 0.\label{eqkasnereinstein3}
\end{equation}
\indent By combining Eqs. \eqref{eqkasnereinstein2}, \eqref{eqkasnereinstein3}, and \eqref{eqkasnerscalar}, one obtains
\begin{equation}
    \phi \sim \sqrt{2}c \log r+\phi_{K1}+...\;,
\end{equation}
for some constant $c$. From Eqs. \eqref{eqkasnereinstein2}, \eqref{eqkasnereinstein3}, and \eqref{eqkasnermaxwx}, one obtains
\begin{equation}
    A_x\sim \frac{A_{K1}}{r^2}+A_{K2}+...\;.
\end{equation}
Thus, one has the following solutions
\begin{align}
    \chi&\sim 2c^2\log r-2A_{K1}^2r^{-2}+\chi_{K1}+...\;, \\ f&\sim -f_{K1}r^{3+c^2}\exp\left(-\frac{A_{K1}^2}{2r^2}\right)+...\;.
\end{align} 
From $\chi\sim 2c^2\log r-2A_{K1}^2r^{-2}+\chi_{K1}$ and $A_x\sim \frac{A_{K1}}{r^2}$, Eqs. \eqref{eqkasnereinstein1} and \eqref{eqkasnermaxwt} can be further simplified as follows
\begin{align}
    N''+N'\left(\frac{c^2}{r}+\frac{A_{K1}^2}{r^3}-\frac{2}{r}\right)-\frac{4NA_{K1}}{r^4}-\frac{2A_{t}'A_{K1}}{r}&=0 ,\\ A_t''+A_t'\left(\frac{c^2}{r}+\frac{A_{K1}^2}{r^3}\right)+\frac{2A_{K1}}{r^3}N\left(\frac{3A_{K1}^2}{r^4}+\frac{c^2}{r}-\frac{3}{r}\right)+\frac{2A_{K1}}{r^3}N'&=0.
\end{align}
When $r\rightarrow\infty$, the equations become
\begin{align}
    A_t''+\frac{A_t'c^2}{r}&=0,\\
    N''+N'\left(\frac{c^2}{r}-\frac{2}{r}\right)-\frac{2A_t'A_{K1}}{r}&=0.
\end{align}
These equations give us
\begin{align}
    A_t &\sim A_{tK1}+\frac{A_{tK2}r^{1-c^2}}{1-c^2}+...\;,\\
    N&\sim N_{K1}+\frac{N_{K2} r^{3-c^2}}{3-c^2}-\frac{N_{K3}A_{K1}r^{1-c^2}}{1-c^2}+...\;.
\end{align}
\indent We can see that, similar to the non-rotating charged black hole in AdS$_4$, the fields generically diverge near the singularity, except for $A_t,A_x,N$ when $c^2>3$. By performing a coordinate transformation $r=\tau^{-2/(3+c^2)}$, the metric and the scalar field become
\begin{align}
    ds^2=&-d\tau^2+a_t \tau^{2p_t}dt^2+a_x\tau^{2p_x}((N_{K1}dt+dx)^2+dy^2),\\
    \phi=&-p_\phi\log\tau+a_\phi,
\end{align}
where $a_t$, $a_x$, and $a_\phi$ are some constants and 
\begin{equation}
    p_t=\frac{c^2-1}{3+c^2},\;\;\;p_x=\frac{2}{3+c^2},\;\;\;p_\phi=\frac{2\sqrt{2 }c}{3+c^2},
\end{equation}
are the Kasner exponents. These exponents satisfy the following relation
\begin{equation}
    p_t+2p_x=1,\;\;\;p_t^2+2p_x^2+p_\phi^2=1.
\end{equation}
All of the Kasner exponents only depend on the integration constant of the scalar field near the singularity, $c$.\\
\indent In this work, we only consider the case with $c^2>3$. Therefore, the leading term of the functions $A_t$ and $N$ near the singularity are constants. In this case, we can define a new coordinate $\tilde{\Omega}=N_K t+x$ near the singularity and the Kasner metric becomes similar to the non-rotating case
\begin{align}
    ds^2=&-d\tau^2+a_t \tau^{2p_t}dt^2+a_x\tau^{2p_x}(d\tilde{\Omega}^2+dy^2).
\end{align}
The value of $c^2$ might change to $0<c^2<1$ under the Kasner transition. When $c^2>3$, the Kasner exponent $p_t$ is positive, and all of them satisfy
\begin{equation}
    \frac{2}{3}<p_t<1,\;\;\;0<p_x<\frac{1}{3},\;\;\;|p_\phi|^2<\frac{2}{3}.
\end{equation}
The relation between the Kasner exponent $p_t$ and the dimensionless boundary deformation parameter $\phi_0/T$ for various values of the rotation parameter $N_{H1}$ can be seen in Figure \ref{fig:kasnerNH}.\\
\indent Note that the Kasner geometry in the interior of hairy black holes depends on the value of $c^2$. For example, the function $N(r)$ behaves very differently when $c^2>3$ or $c^2<3$ at large $r$. Furthermore, the function $A_t(r)$ also exhibits similar behavior between $c^2>1$ and $c^2<1$. The value of $c^2$ can change under a mechanism called the Kasner inversion, where
\begin{equation}
    c^2\rightarrow\frac{1}{c^2},\;\;\;\text{at}\;\;\;r\gg r_{\text{inv.}},
\end{equation}
with $r_{\text{inv}.}\gg r_H$ denotes the location of the inversion point. When this happens, the Kasner exponent $p_t$ changes sign and it determines whether the term $\tau^{p_t}$ in the Kasner metric grows or shrinks. One may anticipate that the solution with $c^2<3$ is unstable since the rotation function $N(r)$ quickly diverges at large $r$, and we may expect that such a solution will eventually undergo a transition to $c^2>3$. This is observed in \cite{Gao2024} for rotating hairy black holes in AdS$_3$. \\
%\indent A more detailed analytical and numerical analysis of the Kasner inversion in stationary charged hairy black holes is beyond the scope of this paper, as our primary purpose is to study their chaotic properties. We mainly focus on the region not far from the horizon, albeit in the interior. We will leave the study of the Kasner inversion in this background for future works.
\section{Fast Scrambling from Rotating and Charged Shock Waves}
In this section, we calculate the out-of-time ordered correlator (OTOC) holographically in the charged and rotating hairy black hole background. The OTOC is defined as
\begin{equation}
    OTOC(t)=\langle\hat{V}(0)\hat{W}(t)\hat{V}(0)\hat{W}(t)\rangle,
\end{equation}
and it vanishes exponentially in time for chaotic systems under small perturbations \cite{jahnke2019,Trunin2021}. In other words, we can write
\begin{equation}
    \frac{\langle\hat{V}(0)\hat{W}(t)\hat{V}(0)\hat{W}(t)\rangle}{\langle\hat{W}\hat{W}\rangle\langle\hat
    {V}\hat{V}\rangle}=1-\varepsilon e^{\lambda_Lt},
\end{equation}
where $\varepsilon\sim\mathcal{O}(1/N)$, is the order of the perturbation and $\lambda_L$ is the Lyapunov exponent. Here, $N$ represents the number of degrees of freedom. This OTOC vanishes at time order $t_*\sim \frac{1}{\lambda_L}\log N$ defined as the scrambling time. We calculate the OTOC of a CFT$_3$, which is dual to a charged and rotating hairy black hole in the AdS$_4$ bulk, holographically. The expectation value $\langle\cdot\rangle$ here is the expectation value with respect to the thermofield double state that is dual to the two-sided eternal black hole spacetime in the bulk \cite{Maldacena2003eternal}.\\
\indent We consider a thermofield double state $|TFD\rangle$ which is perturbed by $\hat{W}(t)$. The OTOC can be represented by correlations between operators in the left and right asymptotic boundaries, $\langle\hat{\mathcal{O}}_A\hat{\mathcal{O}}_B\rangle_W$, where $\langle\cdot\rangle_W$ is the expectation value with respect to the perturbed thermofield double state by the operator $\hat{W}(t)$. The vanishing of this correlator is related to the vanishing of the mutual information
\begin{equation}
    I(A:B)=S_A+S_B-S_{A\cup B},
\end{equation}
through inequality relation \cite{Wolf2008}
\begin{equation}
    I(A:B)\geq\frac{(\langle\hat{\mathcal{O}}_A\hat{\mathcal{O}}_B\rangle-\langle\hat{\mathcal{O}}_A\rangle\langle\hat{
    \mathcal{O}}_B\rangle)^2}{2\|\hat{\mathcal{O}}_A\|^2\|\hat{\mathcal{O}}_B\|^2},
\end{equation}
where $\|\cdot\|$ denotes the operator norm. Here, both $A$ and $B$ are subregions in the asymptotic boundary. The scrambling time $t_*$ can be obtained from the time scale when the mutual information $I(A:B)$ vanishes. \\
\indent In this work, we calculate the entanglement entropy $S_A,S_B,S_{A\cup B}$ holographically using the Ryu-Takayanagi holographic entanglement entropy formula \cite{Ryu2006PRL,Ryu2006JHEP}
\begin{equation}
    S_A=\frac{\mathcal{A}(\gamma_A)}{4G_N},
\end{equation}
where $\mathcal{A}(\gamma_A)$ denotes the area of a minimal surface $\gamma_A$ in the AdS bulk that follows homology condition $\partial\gamma_A=\partial A$. We consider regions $A$ and $B$ that are located on the left and right asymptotic boundaries, respectively. The entanglement entropy $S_{A \cup B}$ is computed from the minimal surface connecting the two boundaries, $\gamma_{A\cup B}$, analogous to the Hartman-Maldacena surface \cite{Hartman2013}. This surface penetrates the black hole interior and is influenced by Dray-'t Hooft gravitational shock waves. Initially, $\mathcal{A}(\gamma_{A\cup B})< \mathcal{A}(\gamma_A)+\mathcal{A}(\gamma_B)$, yielding $I(A:B)>0$. After the shock waves enter the black hole, $\mathcal{A}(\gamma_{A\cup B})$ becomes longer while $\mathcal{A}(\gamma_A)$ and $\mathcal{A}(\gamma_B)$ are unaffected and remain constant. Therefore, a transition from $S_{A\cup B}=\frac{\mathcal{A}(\gamma_{A\cup B})}{4G_N}$ to $S_{A\cup B}=S_A+S_B$ may occur when $\mathcal{A}(\gamma_{A\cup B})=\mathcal{A}(\gamma_A)+\mathcal{A}(\gamma_B)$, at which point the mutual information between $A$ and $B$ drops to zero.
\\
\indent The first part of this section calculates the metric solution in the Kruskal coordinates when the black hole background is perturbed by rotating and charged gravitational shock waves. The shock waves come from the left boundary at the insertion time $\tau_w$ and goes through the horizon, carrying angular momentum $\mathcal{L}$ and charge $\mathcal{Q}$. The second part calculates the minimal surface that stretches from the left to the right asymptotic boundaries. This minimal surface depends on the shock waves' insertion time $\tau_w$. When $\tau_w$ is large enough, the mutual information vanishes. In the third part, we obtain the black hole's chaotic data, including the Lyapunov exponent and the scrambling time delay, from the mutual information. We see how these chaotic aspects of the black hole are influenced by the boundary deformation $\phi_0$ and how they are related to the Kasner exponents in the interior.
\subsection{Metric in Rotating Shock Waves Background}
\begin{figure}
    \centering
    \includegraphics[width=0.45\linewidth]{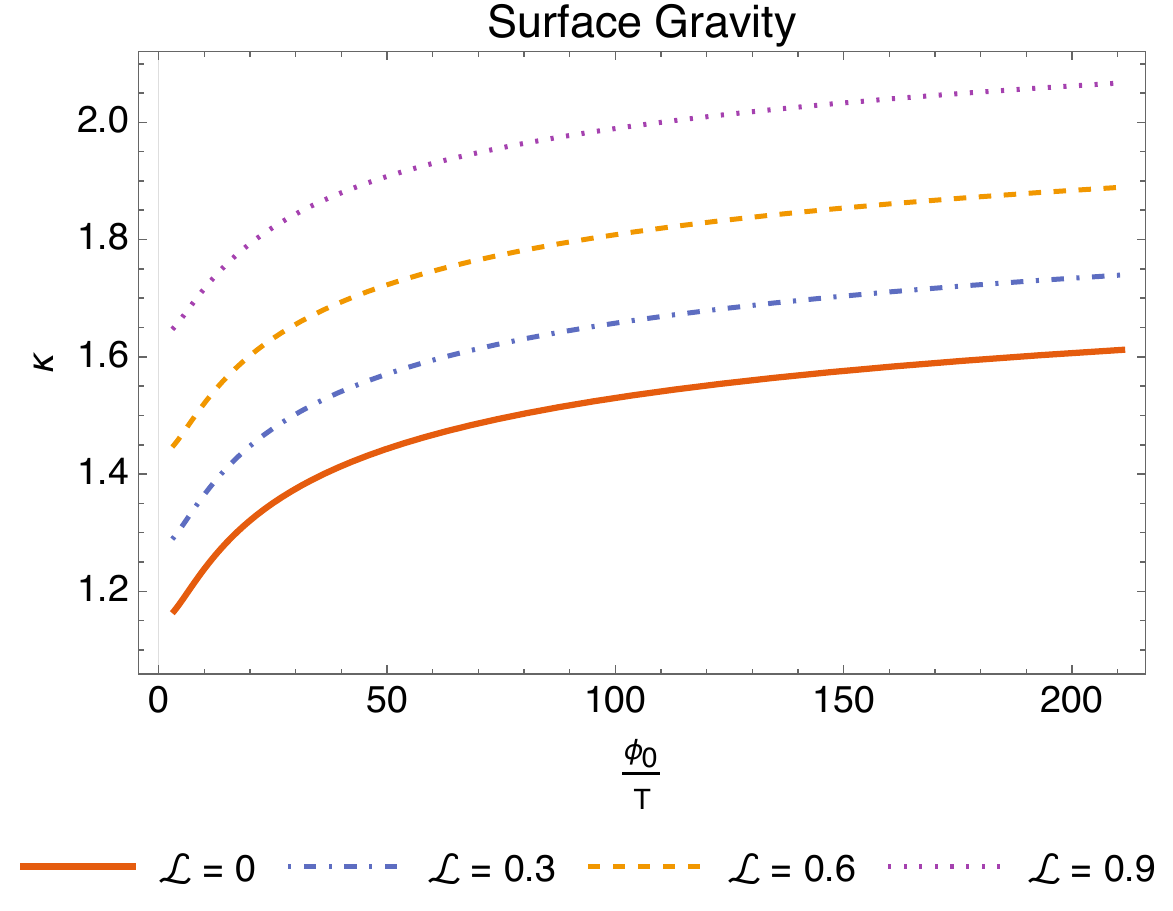}
    \includegraphics[width=0.40\linewidth]{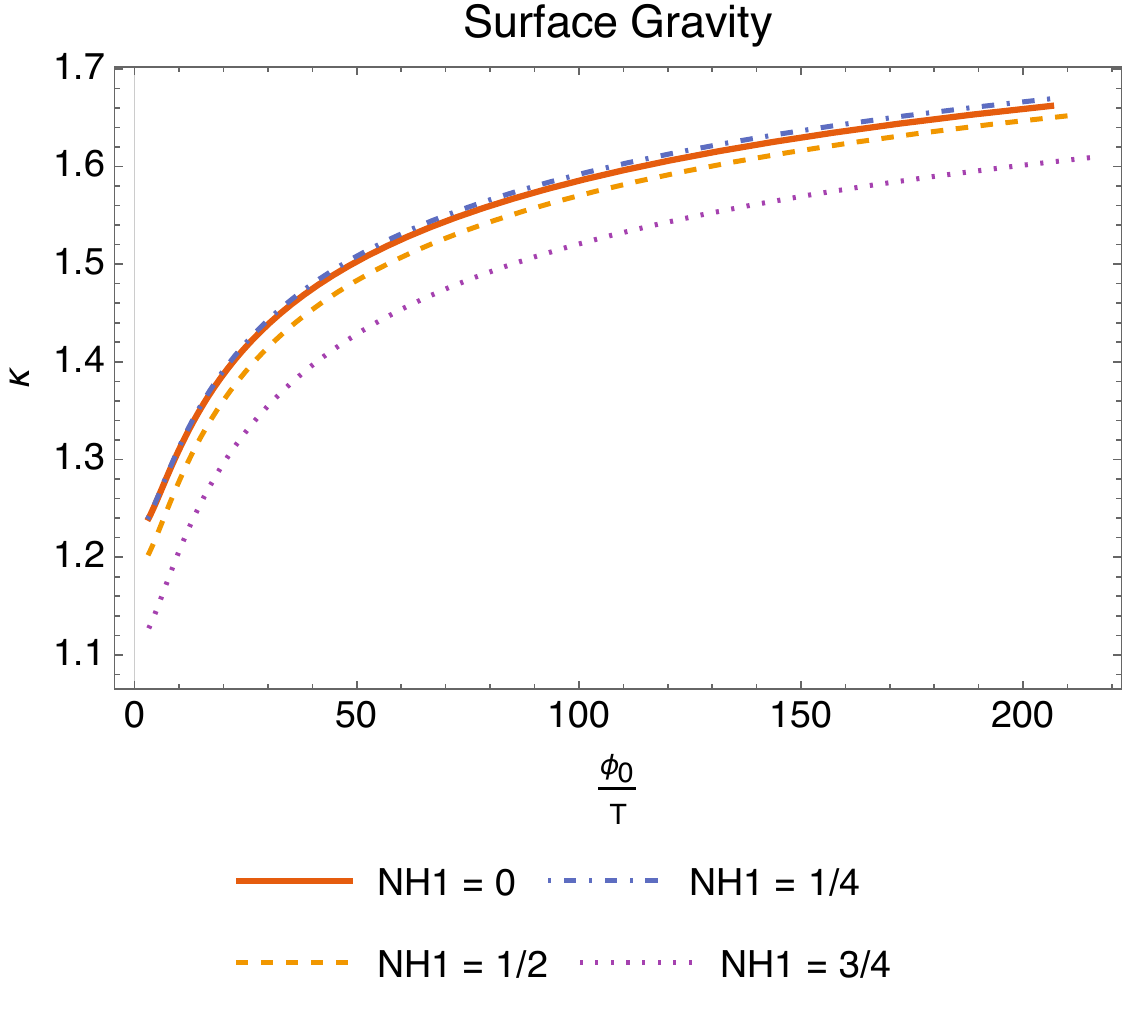}
    \caption{Surface gravity $\kappa=\frac{2\pi T_H}{1-\Omega_H\mathcal{L}}$ versus boundary deformation parameter $\phi_0/T$ when the angular momentum $\mathcal{L}$ and the rotation parameter $N_{H1}$ are varied.}
    \label{fig:surfacegravity}
\end{figure}

In this subsection, we consider a metric in the frame of a rotating shock waves at $y=$ constant traveling at the speed of light. This rotating shock waves have energy $\mathcal{E}$ and angular momentum $\mathcal{L}$. It also follows a null geodesic $\xi^\mu$ which satisfies
\begin{equation}
    \xi\cdot\xi=0,\;\;\;\xi\cdot\zeta_t=-\mathcal{E},\;\;\;\xi\cdot\zeta_x=\mathcal{L},
\end{equation}
where $\zeta_t=\partial_t$ and $\zeta_x=\partial_x$ are the Killing vectors associated with time translational symmetry and axial symmetry due to rotation. The metric at $y=\text{constant}$ is given by
\begin{equation}
    ds^2=\frac{1}{r^2}\bigg(-f(r)e^{-\chi{(r)}}dt^2+\frac{dr^2}{f(r)}+(N(r)dt+dx)^2\bigg).
\end{equation}
At the asymptotic boundary, we have set $N(0)=0$ so that the boundary metric is non-rotating. Using this metric, we can solve the geodesic equations and find the solution (with unit energy $\mathcal{E}=1$):
\begin{align}
    \xi^{(\pm)}=&\mp\frac{r^2(1+\mathcal{L}N)}{fe^{-\chi}}\partial_t\mp\frac{r^2(\mathcal{L}fe^{-\chi}-N(1+\mathcal{L}N))}{fe^{-\chi}}\partial_x\\\nonumber
    &+\sqrt{r^4(-\mathcal{L}^2f+e^{\chi}(1+\mathcal{L}N)^2)}\partial_r.
\end{align}
In this case, $\mathcal{L}$ is the angular momentum per unit energy. The $\xi_\mu^{(-)}$ solution is obtained from reversing the axial symmetry $-\mathcal{E}\rightarrow\mathcal{E}$ and $\mathcal{L}\rightarrow-\mathcal{L}$.\\
\indent Using the metric at $y=$ constant, the geodesic $\xi_{\mu}^{(\pm)}$ can be contracted so that we have
\begin{align}
    \xi^{(\pm)}_\mu dx^\mu=dr_*\pm d\tau,
\end{align}
where
\begin{align}
r_*\equiv\int\frac{\sqrt{e^\chi(1+\mathcal{L}N)^2-\mathcal{L}^2f}}{f}dr=\int\frac{dr}{\tilde{f}},
\end{align}
is the tortoise coordinate with
\begin{equation}\label{ftilde}
    \tilde{f}\equiv\frac{f}{\sqrt{e^{\chi}(1+\mathcal{L}N)^2-\mathcal{L}^2f}},
\end{equation}
and $\tau=t-\mathcal{L}x$. Writing $\xi^{(+)}\cdot dx=du$ and $\xi^{(-)}\cdot dx=dv$, the metric then becomes
\begin{align}
    ds^2&=F(r)\xi^{(+)}_\mu dx^\mu \xi^{(-)}_\nu dx^\nu+h(r)(dx+h_\tau(r)d\tau)^2,\\
    &=F(r)dudv+h(r)(dx+h_\tau(r)d\tau)^2,
\end{align}
where
\begin{equation}
    F(r)=\frac{f(r)e^{-\chi(r)}}{r^2(-\mathcal{L}^2f(r)e^{-\chi(r)}+(1+\mathcal{L}N(r))^2)},\label{F(r)}
\end{equation}
\begin{equation}
    h(r)=\frac{-\mathcal{L}^2f(r)e^{-\chi(r)}+(1+\mathcal{L}N(r))^2}{r^2},\label{h(r)}
\end{equation}
\begin{equation}
    h_\tau(r)=\frac{(\mathcal{L}f(r)e^{-\chi(r)}-N(r)(1+\mathcal{L}N(r)))^2}{(-\mathcal{L}^2f(r)e^{-\chi(r)}+(1+\mathcal{L}N(r))^2)^2}.\label{htau(r)}
\end{equation}
One can see that $h_\tau(r)$ does not vanish at the horizon. If we want to have a function that vanishes at the horizon, we can simply perform a shift $h_\tau(r)\rightarrow h_\tau(r)+\eta_1$, with
\begin{equation}
    \eta_1=-\frac{N(r_H)}{1+\mathcal{L}N(r_H)}=\frac{\Omega_H}{1-\Omega_h\mathcal{L}},
\end{equation}
where $\Omega_H=-N(r_H)$ is the angular velocity of the horizon. We also shift $x\rightarrow \eta_2\tilde{x}+\eta_1\tau$, where
\begin{equation}
    \eta_2=\frac{1}{1+\mathcal{L}N(r_H)}=\frac{1}{1-\Omega_H\mathcal{L}},
\end{equation}
so that the metric now becomes
\begin{equation}
    ds^2=F(r)dudv+\tilde{h}(r)(d\tilde{x}+\tilde{h}_\tau(r)d\tau)^2,
\end{equation}
with $\tilde{h}(r)=\eta_2^2h(r)$ and $\tilde{h}_\tau(r)=\eta_2^{-1}(h_\tau(r)+\eta_1)$.\\ \indent The next step is to transform $u,v$ to $U,V$, where
\begin{equation}
    U=-e^{\kappa u},\;\;\; V=e^{\kappa v},
\end{equation}
with
\begin{equation}
    \kappa=\frac{2\pi T_H}{1-\Omega_H\mathcal{L}}.
\end{equation}
This transformation is different compared to the standard Kruskal-Szekeres coordinates as the surface gravity now picks an extra factor that depends on both $\Omega_H$ and $\mathcal{L}$. This is done so that the coordinates are affine at the horizon, as this is necessary in generating the Dray-'t Hooft solution for the rotating gravitational shock waves \cite{Malvimat2023,Prihadi2023}. In this coordinate system, the metric becomes
\begin{equation}
    ds^2=\frac{F(UV)}{\kappa^2 UV}dUdV+\tilde{h}(UV)\bigg(d\tilde{x}+\frac{\tilde{h}_\tau(UV)}{2\kappa UV}(UdV-VdU)\bigg)^2.
\end{equation}
\indent The parameter $\kappa$ is related to the original horizon's surface gravity $\kappa_0=2\pi T_H$ as
\begin{equation}
    \kappa=\frac{\kappa_0}{1-\Omega_H\mathcal{L}}.
\end{equation}
The value of $\Omega_H\mathcal{L}\leq1$ is satisfied for a geodesic that comes from the asymptotic boundary and going through the horizon \cite{Malvimat2022, Malvimat2023}. Furthermore, the equality $\Omega_H\mathcal{L}=1$ can be achieved in the extreme limit where $T_H\rightarrow 0$. In this limit, $\kappa$ approaches a non-zero value $\kappa_{\text{ext}}$. Previously, we have shown that, in the Kerr-Sen-AdS black hole case \cite{Prihadi2023}, $\kappa_{\text{ext}}$ coincides with the left-moving Frolov-Thorne temperature $T_L$ in the context of the Kerr/CFT correspondence \cite{Sakti2022,Sakti2023}. The modified surface gravity increases with the boundary deformation parameter, as shown in Figure \ref{fig:surfacegravity}.\\
\indent We now add rotating and charged gravitational shock waves in the Dray-'t Hooft solution \cite{Dray1985, DRAY1985173} which gives us a shift in the metric as
\begin{equation}
    ds^2\rightarrow ds^2-\frac{F(UV)}{\kappa^2 UV}\alpha\delta(U)dU^2,
\end{equation}
where the shock waves strength $\alpha$ is given by
\begin{equation}
    \alpha=\frac{\beta_HE_0}{S}(1-\Omega_H\mathcal{L}-\mu_e\mathcal{Q})e^{\frac{2\pi \tau_w}{\beta}}.
\end{equation}
This expression for $\alpha$ has been derived previously in \cite{Prihadi2023} for gravitational shock waves with energy $E_0$, angular momentum per unit energy $\mathcal{L}$, and charge per unit energy $\mathcal{Q}$. Here, $\beta_H=\frac{1}{T_H}$ is the inverse of the Hawking temperature, $S$ is the Bekenstein-Hawking entropy, $\tau_w$ is the insertion time of the shock waves, $\Omega_H$ is the angular velocity of the black hole horizon obtained from $N(r_H)=-\Omega_H$, and $\mu_e$ is the electric chemical potential obtained from $A_{tb0}=\mu_e$. In general, the gravitational shock waves can also contain magnetic charge $\mathcal{P}$ so that $\alpha$ depends on $\mu_m\mathcal{P}$ as well. However, we only consider the case where $\mathcal{P}=0$ for simplicity.\\
\indent The gravitational shock waves create a shift in the $V$ direction of the Kruskal-Szekeres coordinates so that
\begin{equation}
    V\rightarrow V+\alpha\Theta(U-U_0),
\end{equation}
where $U_0$ is the point where the shock waves are inserted. This shift extends the wormhole geometry and, as we will show later, increases the length of the minimal surface connecting the left and right boundaries. Even though we consider the energy of the perturbation in the order of a few Hawking quanta, or $\beta_H E_0\sim 1$, as $S\rightarrow \infty$ (large black hole's degrees of freedom) this $\alpha$ parameter can be of order $\alpha\sim\mathcal{O}(1)$ when the insertion time is large, i.e. $t_w\sim\log S$. At this time scale, the correlation between the two asymptotic boundaries vanishes, signaling the onset of scrambling.
\subsection{Calculations of the Minimal Surface}
\begin{figure}
    \centering
    \includegraphics[width=0.90\linewidth]{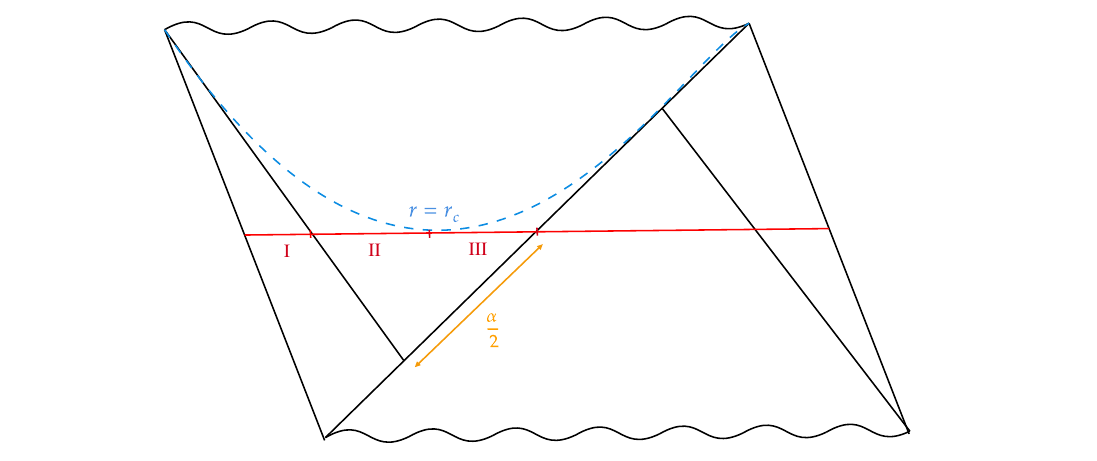}
    \caption{Penrose diagram of the black hole under shock waves geometry background. The minimal surface is denoted by the red line, which is further divided into three segments: I, II, and III. Note that the length of segment II and segment III is equal. The critical value of $r$ is denoted by $r_c$.}
    \label{fig:penrosediagram}
\end{figure}

Next, we compute the disruption of the mutual information between the two TFD states living on the left and right Kruskal boundaries. In particular, we take a constant $y$ slice and consider two copies of CFTs, which we call subsystems $A$ and $B$ living on the left and right boundaries, respectively. The induced metric ($ds^2$) can be expressed in terms of the coordinates $\{r, \tau, x\}$.
\begin{equation}
    ds^2 = F\left(-d\tau^2 + \frac{dr^2}{\tilde{f}^2}\right) + \tilde{h}(d\tilde{x}+\tilde{h}_\tau d\tau)^2,
    \label{eq:metric}
\end{equation}
where $\tilde{f}$ is defined in Eq. \eqref{ftilde}.\\ \indent The area of the extremal surface ($\mathcal{A}$) can be obtained by extremizing the following functional.
\begin{equation}
    \mathcal{A} = S_x \int d\tau \sqrt{h} \sqrt{-F + F\tilde{f}^{-2}\dot{r}^2},
    \label{eq:area_functional}
\end{equation}
where $\dot{r}=dr/d\tau$ and $S_x$ is the length of the $\tilde{x}$ coordinate with $S_x=2\pi$ if we identify $\tilde{x}\sim \tilde{x}+2\pi$. This leads to a conserved quantity $\gamma$, which arises from the fact that the area functional is independent of the $\tau$ coordinate, and can be calculated by solving the EoM.
\begin{equation}
    \gamma = \frac{-F\sqrt{\tilde{h}}}{\sqrt{-F+F\tilde{f}^{-2}\dot{r}^2}} = \sqrt{-F_0\tilde{h}_0}.
    \label{eq:gamma_conserved}
\end{equation}
In the expression above, $F_0$ and $\tilde{h}_0$ are the values of functions $f$ and $\tilde{h}$ at a point where $\dot{r}=0$ defined as the turning point.\\
\indent The coordinate $\tau$ can be expressed as a function of $r$ by inverting the conservation equation in Eq. \eqref{eq:gamma_conserved}.
\begin{equation}
    \tau(r) = \int \frac{dr}{\tilde{f}\sqrt{1+\gamma^{-2}F\tilde{h}}}.
    \label{eq:tau_of_r}
\end{equation}
We can define the tortoise coordinate $r_*$ as a function of the radial coordinate $r$.
\begin{equation}
    r_*(r) = -\int_\infty^r \frac{dr'}{\tilde{f}(r')}.
    \label{eq:tortoise}
\end{equation}
The extremal surface is then analyzed using the null coordinates \(U\) (outgoing) and \(V\) (ingoing). In this subsection, we follow the discussions made by \cite{Leichenauer_2014,Malvimat2022} in which we compute the area of the extremal surface by dividing the extremal surface into three different segments as shown in Figure \ref{fig:penrosediagram}. As one takes the limit $\alpha \to 0$, one will recover the usual Kruskal geometry where the coordinate $V$ merge into one continuous straight line.\\
\indent First, the segment I is taken from the left boundary $(U,V)=(1,-1)$ to $(U,V)=(U_1,0)$. Along the first segment starting from the horizon, the outgoing null coordinate \(U_1^2\) is found by evaluating the integral in Eq. \eqref{eq:segment1}. 
\begin{equation}
    U_1^2 = \exp\left[-2\kappa \int_{r_+}^r \frac{dr}{\tilde{f}}\left(1-\frac{1}{\sqrt{1+\gamma^{-2}F\tilde{h}}}\right)\right] \label{eq:segment1},
\end{equation}
The second segment is then found by taking $(U,V)=(U_1,0)$ to $(U,V)=(U_2,V_2)$. The change in this coordinate as the surface falls along the second segment from its turning point \(r_0\) back toward the horizon is given by the ratio in Eq. \eqref{eq:segment2}. By combining these, the total value of the null coordinate \(U_2^2\) at the end of the second segment can be expressed as the single integral from the turning point to the boundary shown in Eq. \eqref{eq:U2_final}.
\begin{align}
    \frac{U_2^2}{U_1^2} = \exp\left[-2\kappa \int_{r_0}^{r_+} \frac{dr}{-f}\left(\frac{1}{\sqrt{1+\gamma^{-2}Fh}}-1\right)\right], \label{eq:segment2} \\
    U_2^2 = \exp\left[-2\kappa \int_{r_0}^{\infty} \frac{dr}{\tilde{f}}\left(1-\frac{1}{\sqrt{1+\gamma^{-2}F\tilde{h}}}\right)\right]. \label{eq:U2_final}
\end{align}
\indent The corresponding null coordinate of the input, \(V_2\), is then determined by relating it to \(U_2\) and a reference surface at \(\tilde{r}\), as shown in Eq. \eqref{eq:V2}. The third and final segment is taken from $(U,V)=(U_2,V_2)$ to $(U,V)=(0,\alpha/2)$ in which we able to relate the parameter \(\alpha\) to \(V_2\),  in Eq. \eqref{eq:segment3} that the dynamics along this final segment are related to the ratio of the changes along the first two. From this, we have
\begin{equation}
    V_2 = \frac{1}{U_2} \exp\left[-2\kappa \int_{\tilde{r}}^{r_0} \frac{dr}{\tilde{f}}\right], \label{eq:V2}
\end{equation}
\begin{equation}
    \frac{\alpha^2}{4V_2^2} = \frac{U_1^2}{U_2^2} .\label{eq:segment3}
\end{equation}
By combining all of these previous derived equations, the final result for \(\alpha\) can be simply expressed as
\begin{equation}
    \alpha = 2\exp(Q_1+Q_2+Q_3), \label{eq:alpha_final}
\end{equation}
i.e., the exponential of the sum of three distinct integrals, \(Q_1\), \(Q_2\), and \(Q_3\) which are explicitly defined as
\begin{align}
    Q_1 &= -2\kappa \int_{\tilde{r}}^{r_0} \frac{dr}{\tilde{f}}, \label{eq:Q1} \\
    Q_2 &= \kappa \int_{r_0}^{\infty} \frac{dr}{\tilde{f}}\left(1-\frac{1}{\sqrt{1+\gamma^{-2}F\tilde{h}}}\right), \label{eq:Q2} \\
    Q_3 &= 2\kappa \int_{r_0}^{r_+} \frac{dr}{-\tilde{f}}\left(\frac{1}{\sqrt{1+\gamma^{-2}F\tilde{h}}}-1\right). \label{eq:Q3}
\end{align}
\indent From Eq. \eqref{eq:Q3}, one can take the limit where $r_0\to r_c$ such that one will obtain the following expression.
\begin{equation}
    \tilde{h}(r_c)F'(r_c)+\tilde{h}'(r_c)F(r_c) = 0.
    \label{eq:divergence_condition}
\end{equation}
Then, we can express the area of the extremal surface using the conserved quantity $\gamma$. The integration is now performed with respect to the radial coordinate $r$ instead of $\tau$.
\begin{equation}
    \mathcal{A} = 2\pi \int \frac{dr}{\tilde{f}} \frac{F\tilde{h}/\gamma}{\sqrt{1+\gamma^{-2}F\tilde{h}}}.
    \label{eq:area_simplified}
\end{equation}
\indent This equation gives the approximate area of the total extremal surface ($\mathcal{A}_{A \cup B}$) in the regime where the parameter $\alpha$ is large. In this limit, the area is dominated by the contribution of the integral $Q_3$ and is shown to be proportional to $\log \alpha$. The subscript `c' denotes the evaluation at the critical radius $r_c$. 
\begin{equation}
    \mathcal{A}_{A \cup B} \approx \frac{4\pi}{\kappa}\sqrt{-F(r_c) \tilde{h}(r_c)} \, Q_3 \approx \frac{4\pi}{\kappa}\sqrt{-F(r_c) \tilde{h}(r_c)} \, \log \alpha.
    \label{eq:area_approx}
\end{equation}
This expression shows that the area growth of the extremal surface is controlled by the factor $\sqrt{-F(r_c) \tilde{h}(r_c)}$. The entanglement entropy is proportional to the area of the region $\mathcal{A}_{A \cup B}$. This area is equal to four times the previously calculated area $\mathcal{A}$, divided by $4G_N$. Hence, $ I(A;B)$ can be obtained as follows.
\begin{align}\label{mutualinformation}
    I(A;B) &= S_A + S_B - S_{A \cup B} \\
&= \frac{\mathcal{A}_A + \mathcal{A}_B}{4G_N} - \frac{\pi}{G_N}\cdot  \sqrt{-F(r_c)\tilde{h}(r_c)} \tau_w \label{eq:mutual_information} \\
&+ \frac{\pi}{\kappa G_N} \sqrt{-F(r_c)\tilde{h}(r_c)} \log S \nonumber \\
&+ \frac{\pi}{\kappa G_N} \sqrt{-F(r_c)\tilde{h}(r_c)} \log \frac{1}{1 - \Omega_H\mathcal{L}-\mu_e\mathcal{Q}}.\nonumber 
\end{align}
When deriving Eq. \eqref{eq:mutual_information}, we assume that the energy of the perturbation, $E_0$, is in the order of a few Hawking quanta, or $\beta_H E_0\sim 1$, as explained previously. It should be noted that this derivation uses the large-$\alpha$ approximation.

\subsection{Chaotic Parameters vs. Boundary Deformation}
\subsubsection{Lyapunov Exponent}
\begin{figure}
    \centering
    \includegraphics[width=0.5\linewidth]{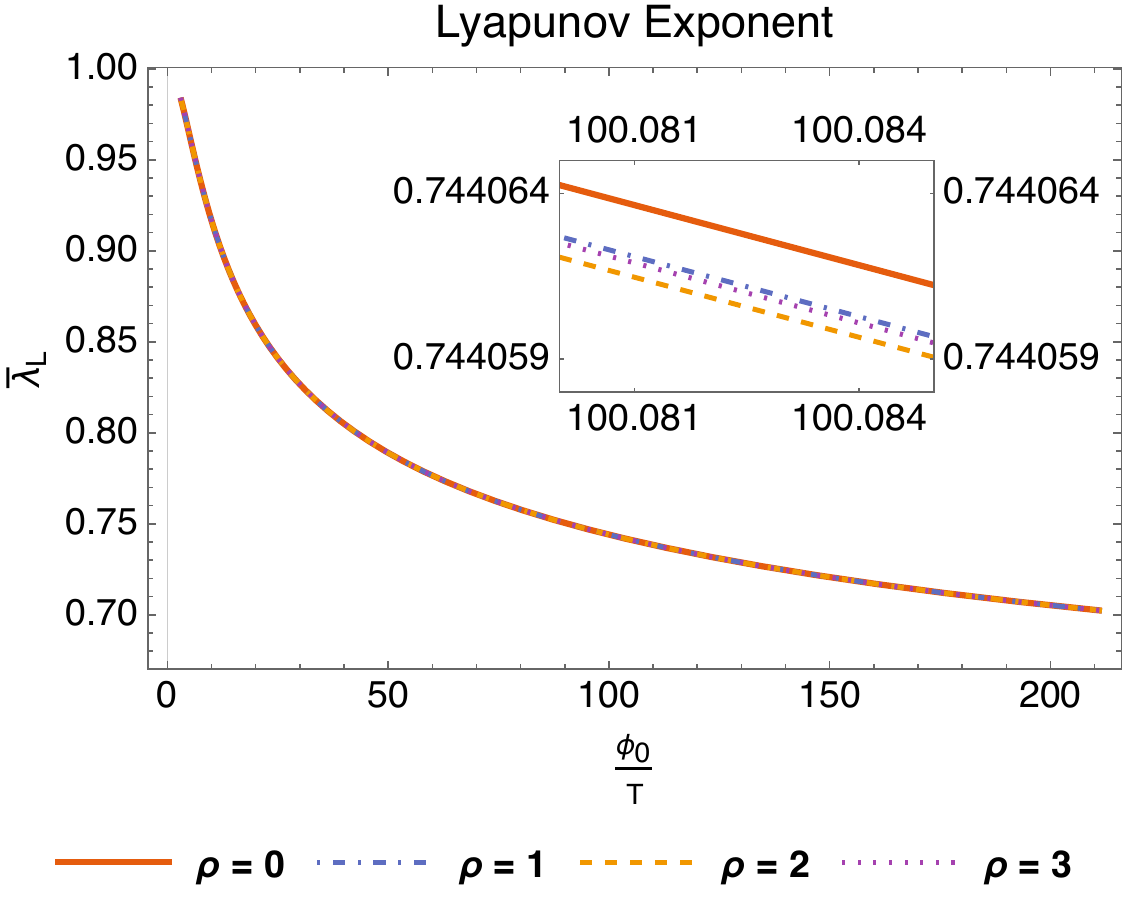}
    \caption{The plot of $\bar{\lambda}_L$ vs $\phi_0/T$ as $\rho$ is varied.}
    \label{fig:LKDefRho}
\end{figure}

The area $\mathcal{A}_{A\cup B}$ is related to the instantaneous Lyapunov exponent,
\begin{equation}
    \mathcal{A}_{A\cup B}=\mathcal{A}_{A\cup B}^{(0)}\lambda_L t_w,
\end{equation}
at late times when $t_w\gg\beta$, where $\mathcal{A}_{A\cup B}^{(0)}$ is the unperturbed version of the area \cite{Malvimat2023,Prihadi2023}. Therefore, the minimal instantaneous Lyapunov exponent is proportional to 
\begin{equation}
    \lambda_L\propto\sqrt{-F(r_c)h(r_c)},
\end{equation}
with the proportionality constant depends on the area of the horizon. In our case, the spacetime is generally non-compact in the $y$ direction. If we want to get rid of this unknown proportionality constant, we may calculate the normalized Lyapunov exponent written as
\begin{equation}
    \bar{\lambda}_L=\frac{\sqrt{-F(r_c)\tilde{h}(r_c)}}{\sqrt{-F(r_c)\tilde{h}(r_c)}|_{\phi_0=0}}.
\end{equation}
\begin{figure}
    \centering
    \includegraphics[width=0.45\linewidth]{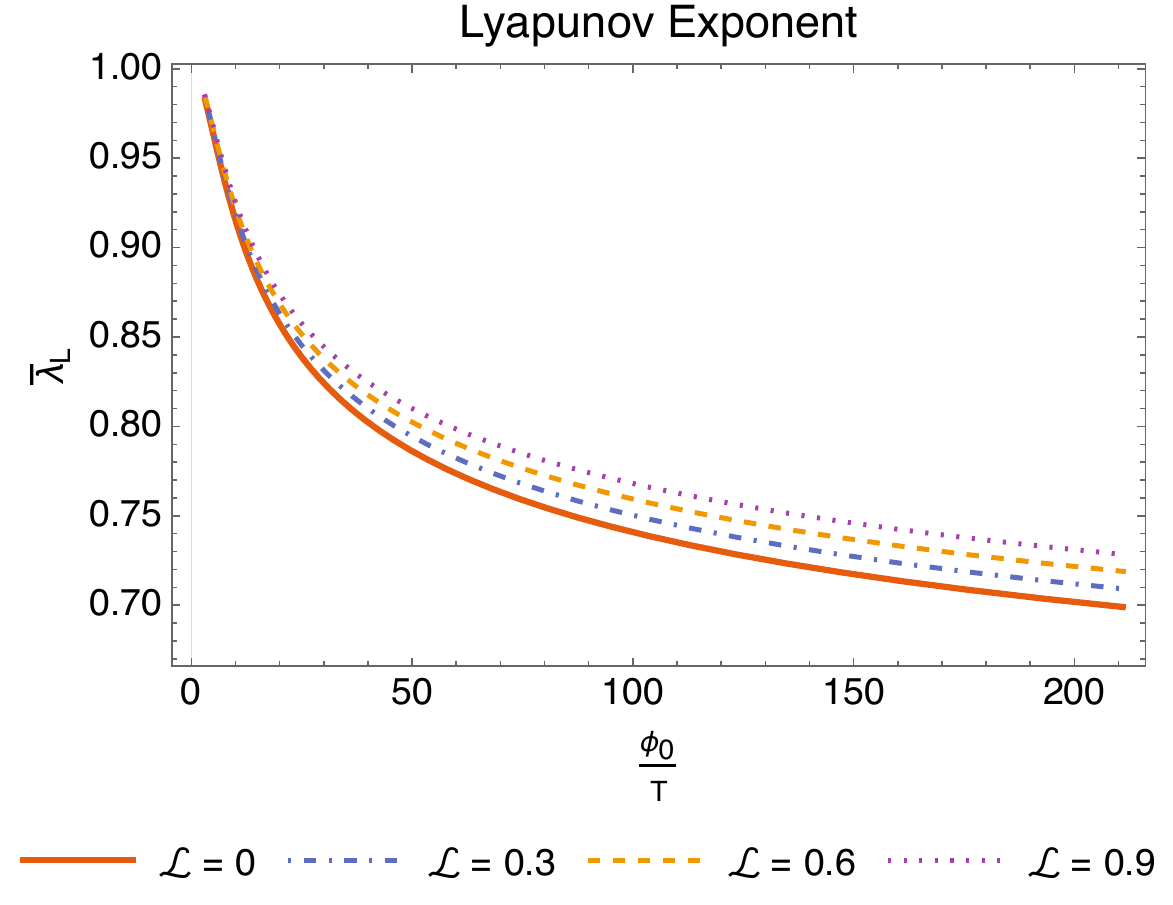}
    \includegraphics[width=0.45\linewidth]{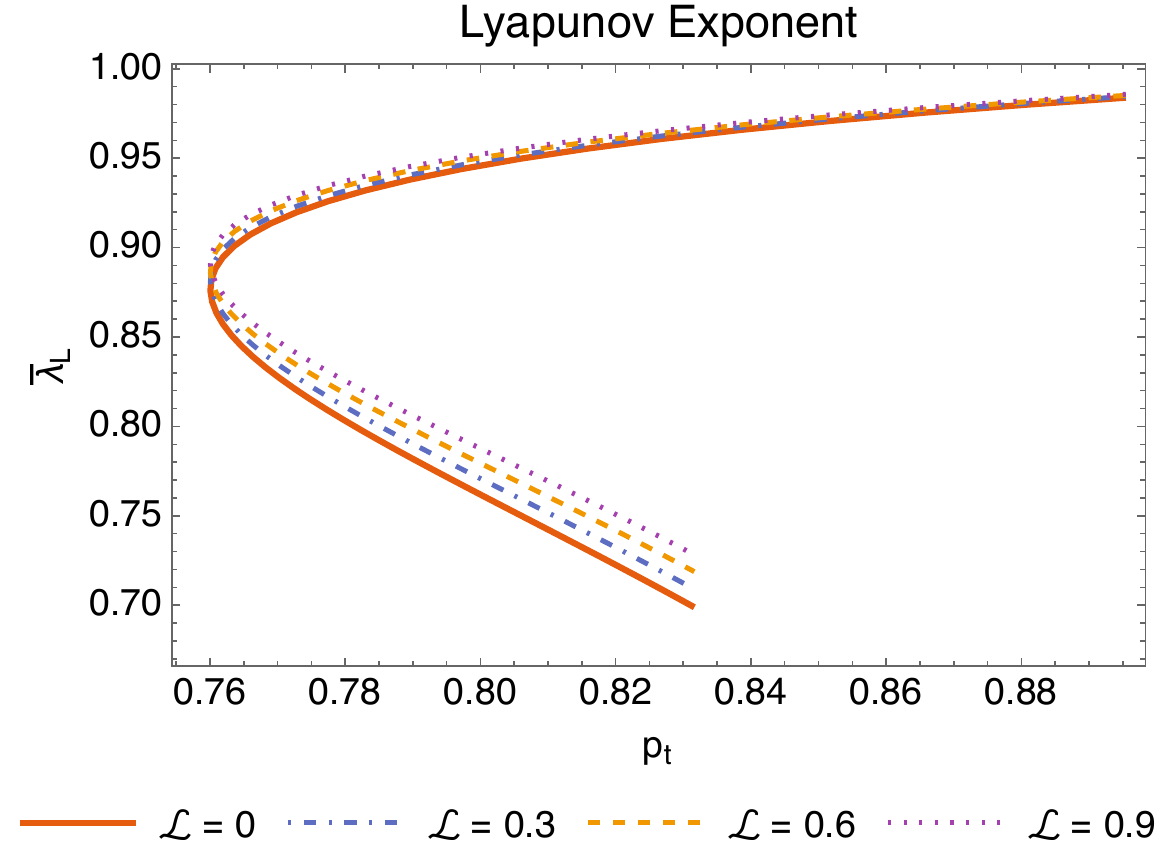}
    \caption{The normalized Lyapunov exponent verus boundary deformation $\phi_0/T$ (left) and versus the Kasner exponent $p_t$ (right) when the angular momentum of the shock waves $\mathcal{L}$ is varied. In this case, we use $N_{H1}=1/2$ and $q=0.10$.}
    \label{fig:LKDefKasnerL}
\end{figure}
\indent As previously shown in \cite{Malvimat2023,Prihadi2023,Prihadi2024}, in general, the minimal instantaneous Lyapunov exponent does not necessarily converge to $\kappa$ in the limit $\phi_0 \rightarrow 0$. Therefore, $\bar{\lambda}_L$ cannot be interpreted as the ratio between the Lyapunov exponent and the surface gravity, as in \cite{prihadi2025scramblingchargedhairyblack}. Nevertheless, $\bar{\lambda}_L$ provides a useful measure to study how $\phi_0$ influences the Lyapunov exponent, indicating whether it increases or decreases in response. We also see how $\mathcal{L}$ affects $\bar{\lambda}_L$, as it depends non-trivially on $\mathcal{L}$ through the definition of the functions $F(r)$ and $\tilde{h}(r)$ in Eqs. \eqref{F(r)} and \eqref{h(r)}.\\
\indent The relation between $\bar{\lambda}_L$ and the deformation parameter $\phi_0/T$ can be seen in Figure \ref{fig:LKDefKasnerL} when we vary the angular momentum $\mathcal{L}$. As we increase the angular momentum, the Lyapunov exponent gets larger. In contrast with the surface gravity shown in Figure \ref{fig:surfacegravity}, $\bar{\lambda}_L$ decreases as the boundary deformation gets stronger. This is expected for the Lyapunov exponent, similar with \cite{prihadi2025scramblingchargedhairyblack}. In $\kappa$, one can directly see that it depends on the scalar field, especially its value at the horizon, from the definition of the hairy black hole's temperature. One can qualitatively see that, as $T\rightarrow 0$, the Lyapunov exponent is expected to vanish as scrambling typically does not occur at black holes with zero temperature \cite{Malvimat2023,Prihadi2023,Prihadi2024}. However, large $\phi_0/T$ does not necessarily imply that the temperature $T$ is approaching to zero. This limit can also be reached at a finite temperature by taking the boundary deformation $\phi_0$ large enough. In this sense, analytical investigations are needed to see the behavior of the Lyapunov exponent with $\phi_0$ at large $\phi_0/T$.\\
\begin{figure}
    \centering
    \includegraphics[width=0.45\linewidth]{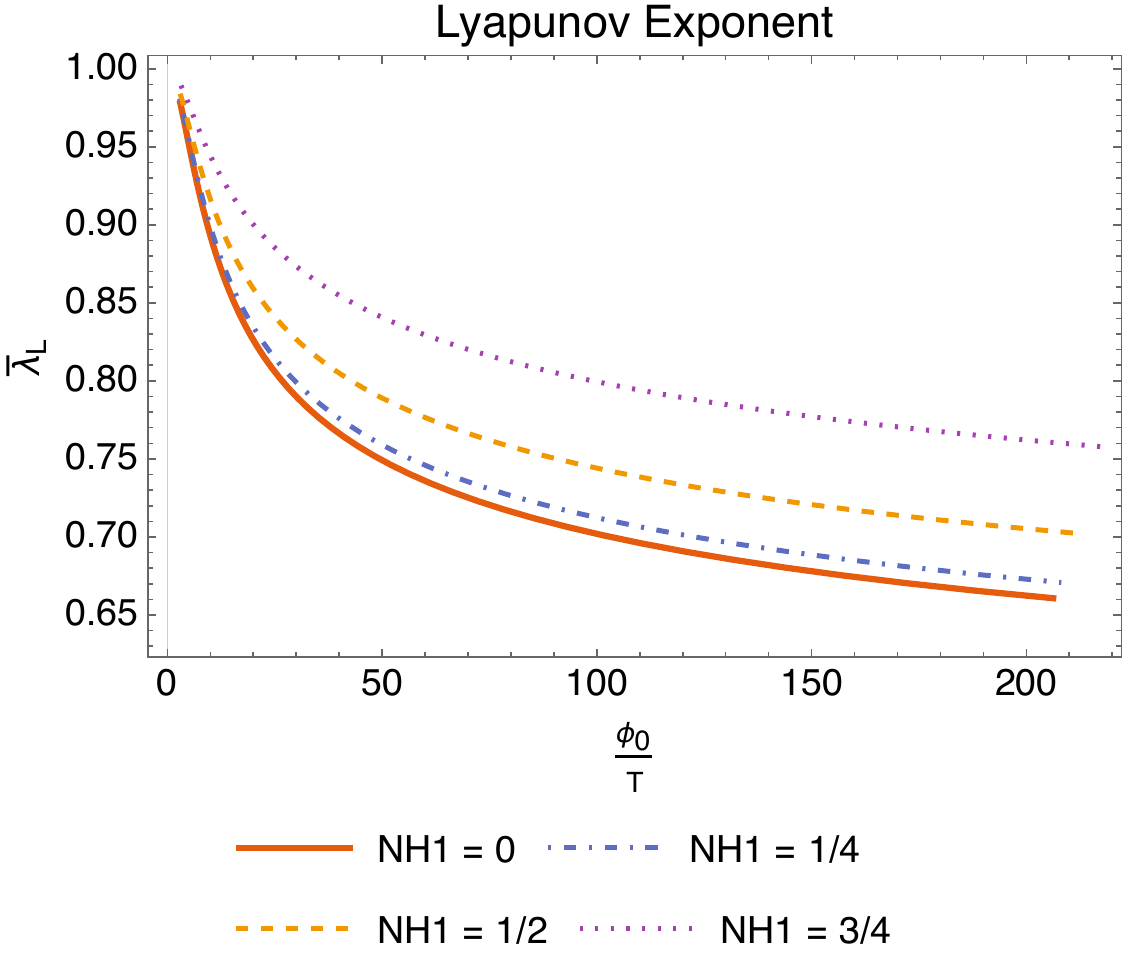}
    \includegraphics[width=0.45\linewidth]{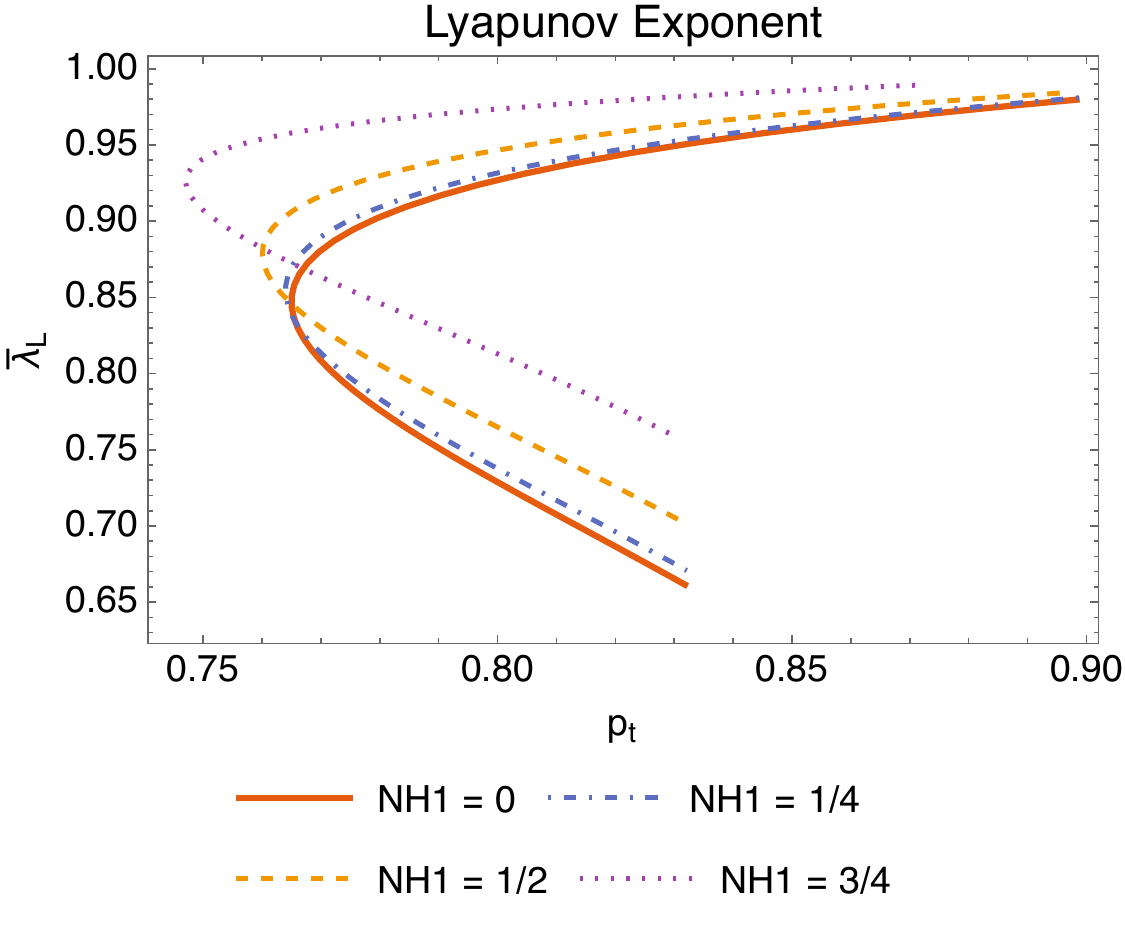}
    \caption{The normalized Lyapunov exponent verus boundary deformation $\phi_0/T$ (left) and versus the Kasner exponent $p_t$ (right) when the rotation parameter $N_{H1}$ is varied.}
    \label{fig:LKDefKasnerNH}
\end{figure}
\indent In contrast with the minimal instantaneous Lyapunov exponent studied in \cite{prihadi2025scramblingchargedhairyblack}, in this case, $\lambda_L$ strongly depends on the angular momentum $\mathcal{L}$ of the shock waves, through the definitions of $F(r)$ and $h(r)$ in Eqs. \eqref{F(r)} and \eqref{h(r)}. The behavior of $\bar{\lambda}_L$ as we vary $\mathcal{L}$ is shown in Figure~\ref{fig:LKDefKasnerL}. The angular momentum $\mathcal{L}$ increases $\bar{\lambda}_L$ when the scalar field is turned on, in agreement to the surface gravity $\kappa$. A similar behavior is also observed in \cite{Prihadi2023}, in the absence of the scalar field. The plot of $\bar{\lambda}_L$ versus the boundary deformation parameter $\phi_0/T$, under of $N_{H1}$, are shown in Figure \ref{fig:LKDefKasnerNH}. We see that the rotation parameter $N_{H1}$ increases the quantum Lyapunov exponent. The charge density $\rho$ does not really affect the Lyapunov exponent, as can be seen in Figure \ref{fig:LKDefRho}.\\
\indent After learning about the relation between Lyapunov exponent $\bar{\lambda}_L$ and the boundary deformation parameter $\phi_0/T$, we may also investigate how $\bar{\lambda}_L$ relates to the Kasner exponents. This can be done through the relation between $\phi_0/T$ and $p_t$ as the boundary deformation controls the value of the Kasner exponents in the interior. This means that, although $\phi_0$ lives in the CFT theory at the boundary, it can tell us about the interior geometry of the black hole. The relations between $\bar{\lambda}_L$ and the Kasner exponent $p_t$ can be seen in Figures \ref{fig:LKDefKasnerL} and \ref{fig:LKDefKasnerNH}, for $\mathcal{L}$, and $N_{H1}$ variations, respectively. We see that the relations are not one-to-one as the relation between $\phi_0/T$ vs $p_t$ in Figure \ref{fig:kasnerNH} are also non-invertible. It is known in holographic theories that properties of the black hole in the bulk cannot solely be determined by non-normalizable modes of the fields which is dual to an operator at the boundary. Additional sub-leading parts of the fields (normalizable modes) are also needed. The reverse is also applicable as one needs subleading corrections to the Kasner geometry to fully determine chaotic data in the boundary \cite{Caceres2024,prihadi2025scramblingchargedhairyblack}.
\subsubsection{Scrambling Time Delay}
\begin{figure}
    \centering
    \includegraphics[width=0.45\linewidth]{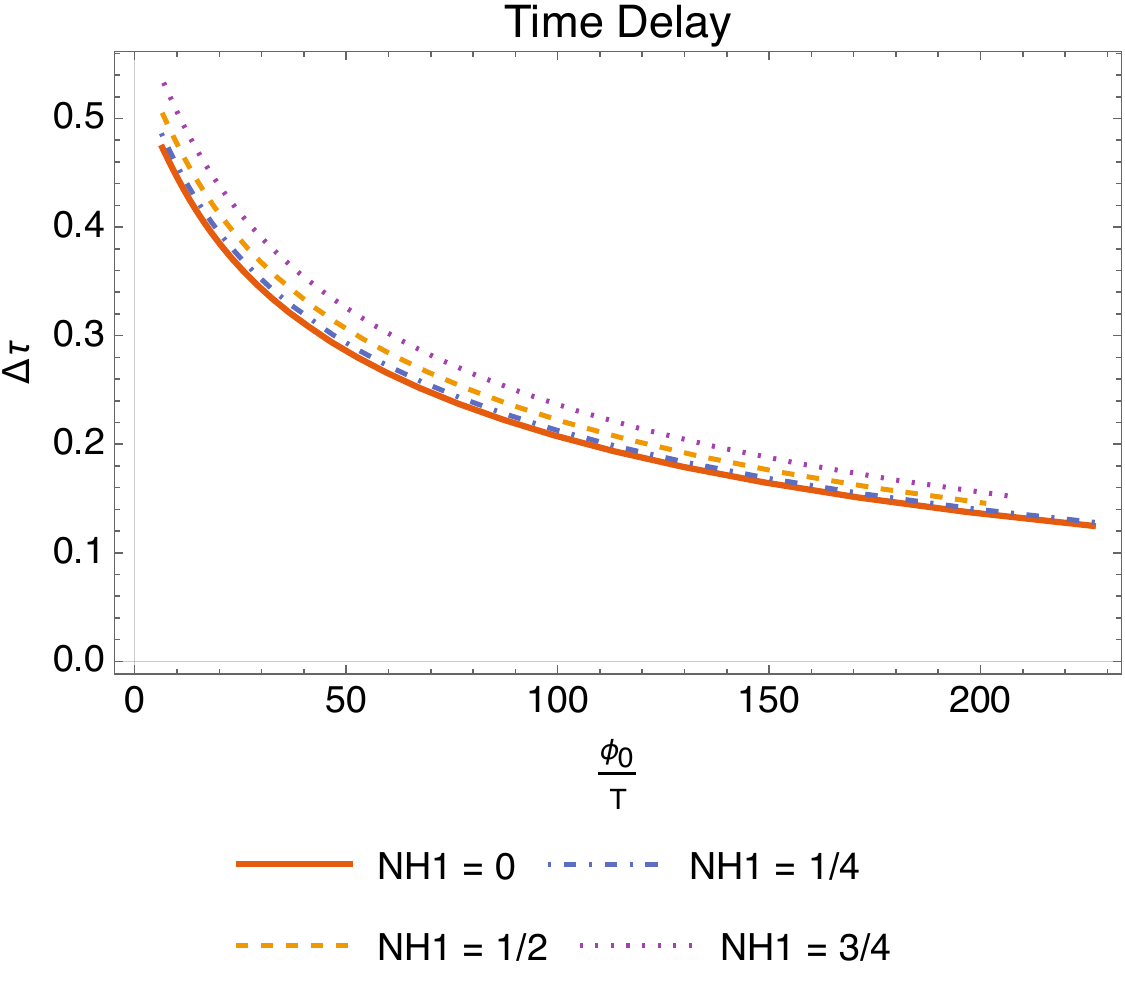}
    \includegraphics[width=0.45\linewidth]{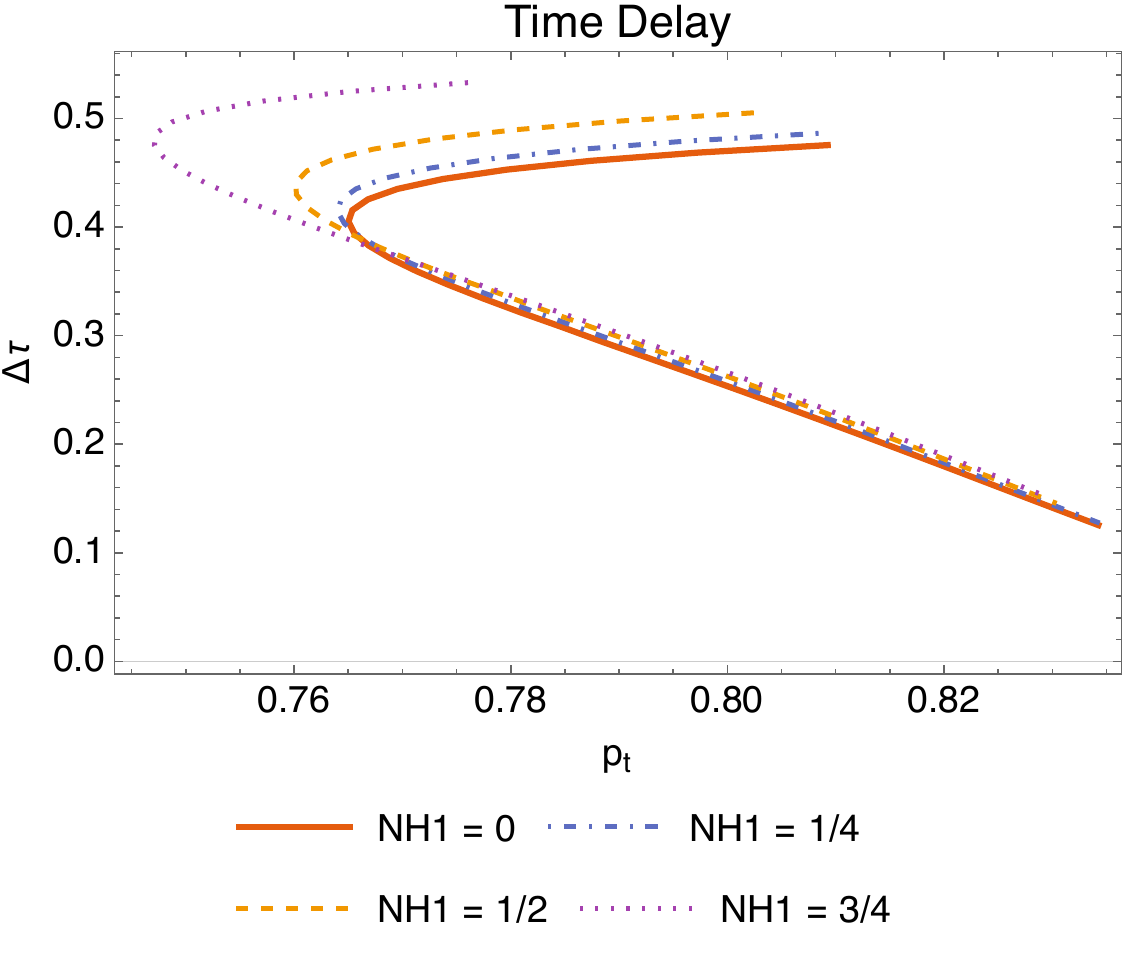}
    \caption{The plot of $\Delta \tau$ versus $\phi_0/T$ (left) and versus $p_t$ (right) when the rotation parameter is varied as $N_{H1}=\{0,1/4,1/2,3/4\}$. In this case, we use $q=0.1,\rho=2,\mathcal{Q}=1/5,\mathcal{L}=0.1$.}
    \label{fig:delayNH}
\end{figure}

\indent The scrambling time $\tau_*$ is defined as the time scale of the insertion time when the mutual information $I(A:B)$ in Eq. \eqref{mutualinformation} vanishes, which is given by
\begin{equation}
    \kappa\tau_*=\log S+\frac{(\mathcal{A}_A+\mathcal{A}_B)}{G_N\sqrt{-F(r_c)\tilde{h}(r_c)}}+\log\bigg(\frac{1-\Omega_H\mathcal{L}}{1-\Omega_H\mathcal{L}-\mu_e\mathcal{Q}}\bigg)+\log\bigg(\frac{1}{1-\Omega_H\mathcal{L}}\bigg).
\end{equation}
The first logarithmic term in black hole entropy $S$ gives us the notion of fast scrambler \cite{Sekino_2008}. The second term does not scale as the black hole entropy and hence it is subleading in the $S\rightarrow \infty$ limit (large black hole's degrees of freedom). The third and fourth terms are also subleading but they appear as extra dynamical terms that is influenced by external probes such as the shock waves' angular momentum $\mathcal{L}$ and charge $\mathcal{Q}$ per unit energy. The third term vanishes when either the black hole or the shock waves are neutral, and it turns out that this term is the scrambling time delay denoted as $\Delta\tau$.\\
\indent The scrambling time delay was first studied in \cite{Horowitz_2022} and generalized to rotating and charged case in \cite{Prihadi2023}. Relation between $\Delta\tau$ and $\phi_0/T$ was first studied in \cite{prihadi2025scramblingchargedhairyblack} for a charged hairy black hole. This notion of delay comes when gravitational shock waves change direction inside the horizon in such a way that the null energy condition is not violated. The time difference between the insertion time and the bounce time represents the delay in the scrambling process. The scrambling time delay for rotating and charged shock waves is given by
\begin{equation}
    \kappa\Delta\tau=\log\bigg(\frac{1-\Omega_H\mathcal{L}}{1-\Omega_H\mathcal{L}-\mu_e\mathcal{Q}}\bigg)+...\;,
\end{equation}
where the $\ldots$ terms are of order thermal time $\mathcal{O}(\beta)$. The value of $\Delta \tau$ is controlled by both the charge $\mathcal{Q}$ and the angular momentum $\mathcal{L}$ of the shock waves. However, the charge plays the dominant role in delaying the onset of scrambling, as $\Delta \tau \rightarrow 0$ when $\mathcal{Q} \rightarrow 0$ or $\mu_e \rightarrow 0$, i.e., when either the charge of the shock waves or that of the black hole vanishes.\\
\begin{figure}
    \centering
    \includegraphics[width=0.45\linewidth]{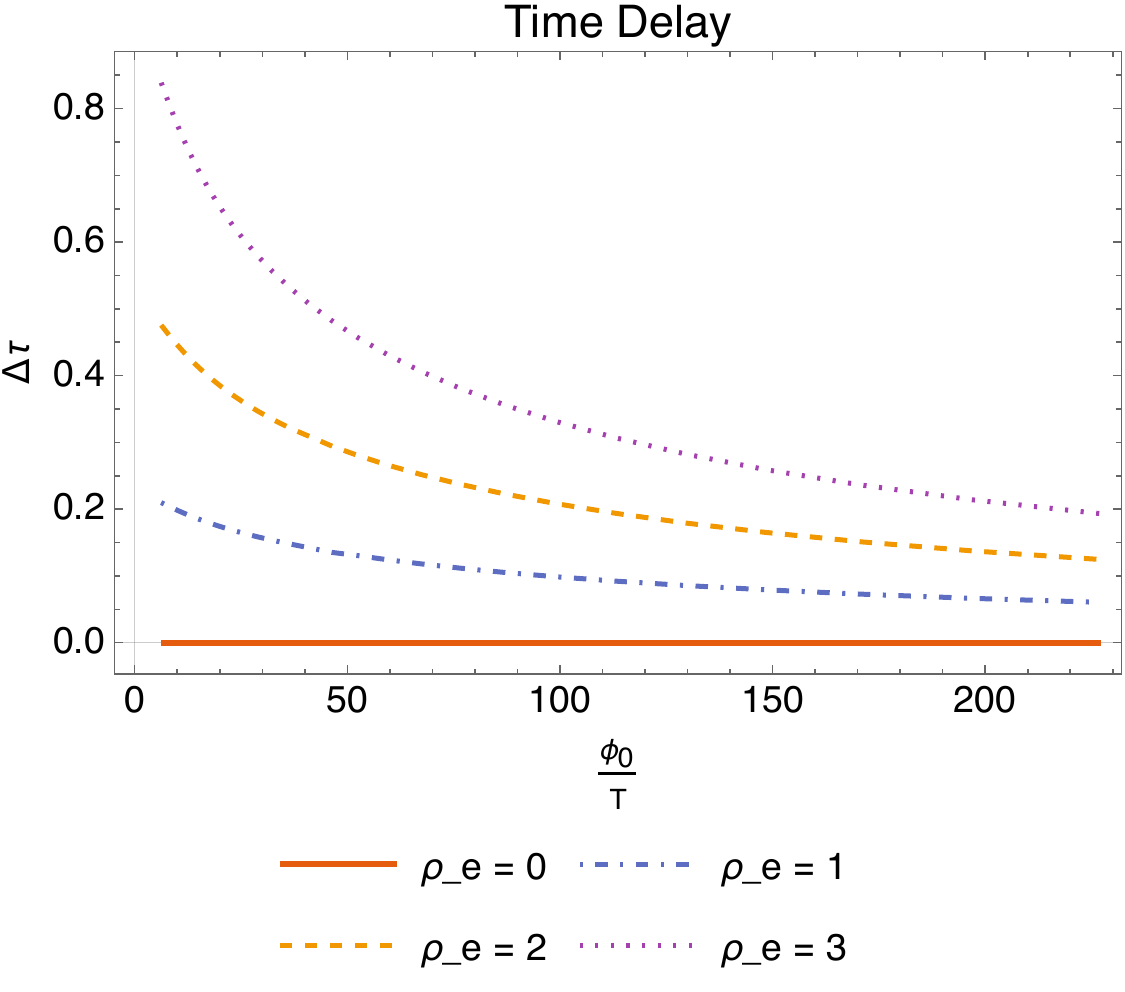}
    \includegraphics[width=0.45\linewidth]{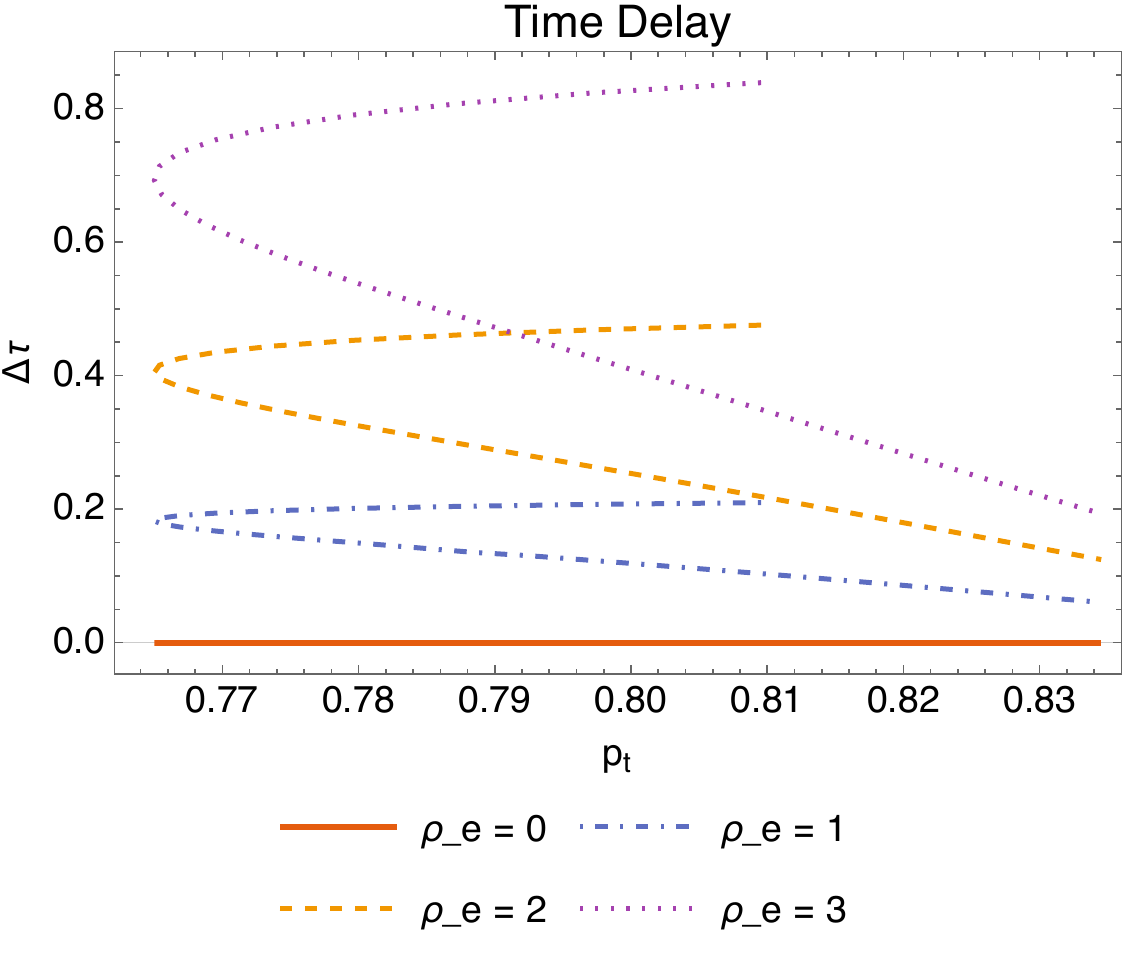}
    \caption{The plot of $\Delta\tau$ vs. $\phi_0/T$ for various $\rho=\{0,1,2,3\}$. Here, we choose $q=0.1,\mathcal{L}=0.1,N_{H1}=-1/2$.}
    \label{fig:delayrho}
\end{figure}
\indent The scrambling time delay is mainly controlled by the black hole's charge density $\rho$ through the chemical potential $\mu_e$. As $\rho$ vanishes, the black hole becomes neutral as the whole Maxwell vector component $A_t(r)$ vanishes. Furthermore, we may expect that, larger black hole's charge leads to larger scrambling time delay as well. This is precisely what is shown in Figure \ref{fig:delayrho} and the behavior is similar to the ones found in the previous work \cite{prihadi2025scramblingchargedhairyblack} without rotation. The strength of the boundary deformation parameter $\phi_0/T$ reduces the scrambling time delay due to the trade-off relation between the scalar field $\phi(r)$ and the Maxwell field $A_t(r)$.\\
\begin{figure}
    \centering
    \includegraphics[width=0.45\linewidth]{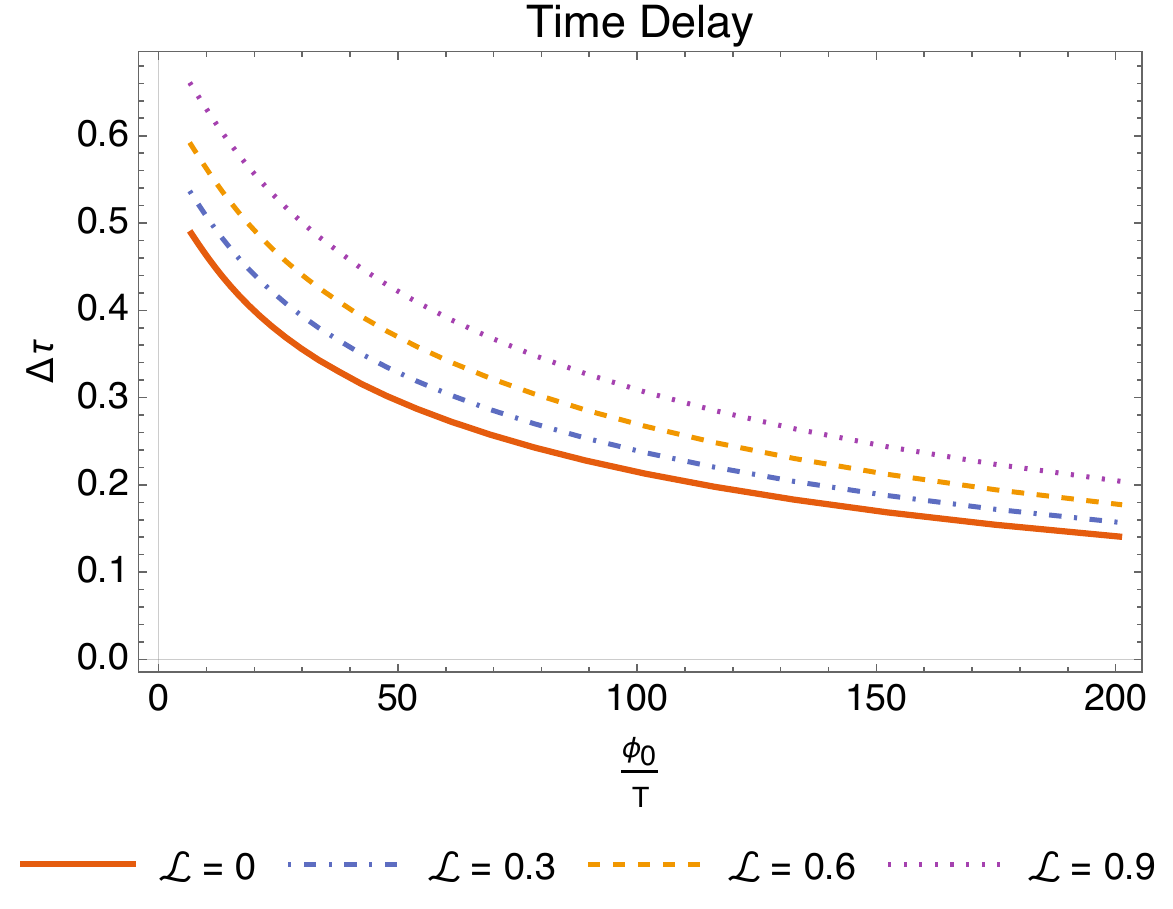}
    \includegraphics[width=0.45\linewidth]{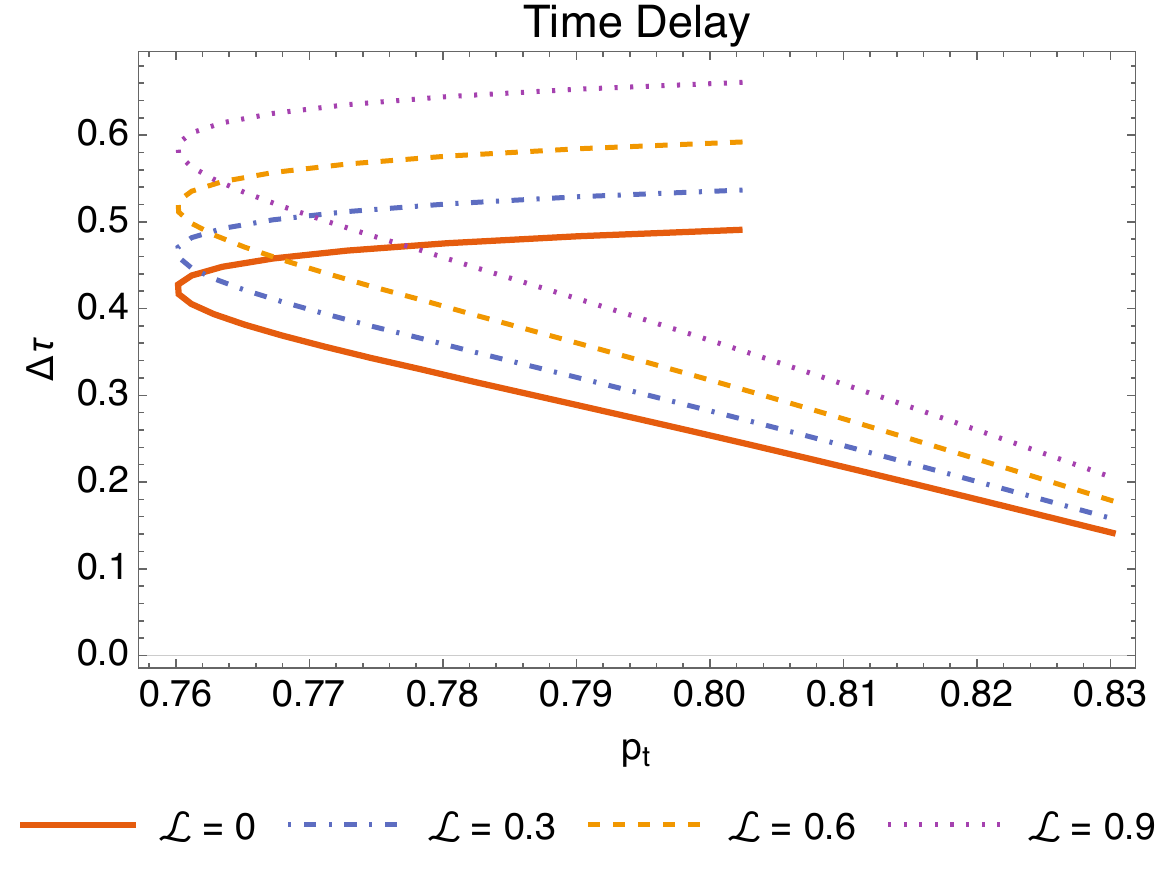}
    \caption{The plot of $\Delta \tau$ versus $\phi_0/T$ (left) and versus $p_t$ (right) when the angular momentum of the shock waves is varied as $\mathcal{L}=\{0,0.3,0.6,0.9\}$. In this case, we use $q=0.1,\rho=2,\mathcal{Q}=1/5,N_{H1}=-1/2$.}
    \label{fig:delayL}
\end{figure}
\indent Other than black hole's charge, $\Delta\tau$ is also controlled by black hole's and shock waves' rotation through $N_{H1}$ and $\mathcal{L}$, respectively. In Figures \ref{fig:delayL} and \ref{fig:delayNH}, one see that both angular momentum of the shock waves $\mathcal{L}$ and the rotation paramter of the black hole $N_{H1}$ prolongs the start of the scrambling process. In the context of the scrambling time delay, both rotation and charge through the terms $\Omega_H\mathcal{L}$ and $\mu_e\mathcal{Q}$, respectively, controls the scrambling time delay in a similar but different way. Furthermore, the relations between the scrambling time delay $\Delta\tau$ and the Kasner exponent $p_t$ are also shown in Figures \ref{fig:delayrho}, \ref{fig:delayL}, \ref{fig:delayNH}.\\
\indent Other models such as in \cite{Sword2022a,prihadi2025scramblingchargedhairyblack} include the Einstein-Maxwell-Scalar coupling term, which gives the action a term
\begin{equation}
    S_{EMS}=\int d^4x\sqrt{|g|}\bigg(-\frac{L^2}{4}\gamma F_{\mu\nu}F^{\mu\nu}|\phi|^2\bigg).
\end{equation}
It turns out that this parameter plays an important role in suppressing the scrambling time delay even more as $\phi_0/T$ becomes large. In this work, we do not see such a suppression behavior from the rotation parameter. We may also expect similar suppression behavior in a stationary charged hairy black hole solution when the model has an EMS coupling term as well. Although we do not consider such a model in this work, this might be interesting to be investigated further.
\section{Summary and Discussions}
In this work, we study the stationary solution to a 4-dimensional asymptotically AdS charged hairy black hole. The real scalar field in the bulk $\phi(r)$ is coupled to the Maxwell field with non-zero components, $A_t(r)$ and $A_x(r)$. We add a function $N(r)$ to the metric, generating a crossing term between $t$ and $x$ coordinates. We then find the solutions numerically, both exterior and interior. We show that the solution does not possess an inner horizon and that an Einstein-Bridge collapse occurs within the interior. Kasner spacetime emerges in the deep interior, with the Kasner exponent controlled by the value of the scalar field at the boundary, $\phi_0$. In this work, we study the Kasner regime where $c^2>3$.\\
\indent In addition to our numerical results regarding the absence of an inner horizon, we note that there are several analytical approaches to demonstrate the nonexistence of an inner horizon in black holes with scalar hair. Arguments based on conserved charges, such as those in \cite{Cai_2021,Hartnoll2021}, apply to hairy black hole solutions with a scalar field directly coupled to a U(1) gauge field $A_\mu$. Other approaches based on the null energy condition (NEC), such as in \cite{An2021}, are applicable to hairy black holes in nonlinear electrodynamics (NED), including those with hyperbolic geometry. However, while these methods work well for static hairy black hole solutions, their extension to stationary black hole solutions is not guaranteed. Moreover, NEC-based analyses cannot be applied to systems with exotic matter \cite{Caldwell_2002,Carroll2003,Huang2008,Prihadi2022Entanglement} or traversable wormholes \cite{Gao_2017,Gao2021SYK,Salcedo2021}, which violate the NEC. At present, a complete proof of the absence of inner horizons in stationary hairy black holes remains an open problem.\\
\indent To study the chaotic properties, rotating and charged shock waves characterized by angular momentum $\mathcal{L}$ and charge $\mathcal{Q}$ are injected into the black hole at the boundary time $\tau_w$ and at fixed coordinate $y=\text{constant}$. We work in Kruskal coordinates representing a stationary observer relative to the rotation of the shock waves. Our analysis demonstrates that the metric functions now depend on the angular momentum $\mathcal{L}$. Using this metric, we calculate a minimal surface that connects the left and right asymptotic boundaries, revealing that it stretches as the insertion time increases. Once the insertion time is sufficiently large, the correlation between the left and right boundaries vanishes.\\
\indent The chaotic aspects of the black hole are obtained from the calculations of the holographic mutual information. The Ryu-Takayanagi surfaces that correspond to the mutual information extend into the interior of the black hole. The minimal surfaces have a turning point in the interior. In this work, we consider a limit when the shock wave parameter $\alpha$ is large. In this limit, the turning point can be obtained by solving Eq. \eqref{eq:divergence_condition}. Therefore, the chaotic properties such as $\lambda_L$ and $\Delta \tau$ now depend on the location of this critical turning point $r_c$ in the interior as $r_c>1$. We then expect that the chaotic properties in the boundary theory are related to the interior solution, such as the Kasner exponent. However, $r_c$ is located near the black hole horizon with $r_c\lesssim 10$. For example, we find that $r_c=1.31725$ when $\phi_0/T=3.39577$. We might expect that the Kasner region after Kasner transition/inversion does not affect the chaotic properties since usually Kasner inversion/transition happens in the deep interior where $r\sim\mathcal{O}(10^x)$ with $x$ is in the order of 10 (see, for example \cite{Sword2022a,Hartnoll2021,Gao2024,Cai2023igv}).\\
\indent The instantaneous quantum Lyapunov exponent is obtained from the rate of the quantum mutual information. Our numerical result shows that the normalized Lyapunov exponent value, $\bar{\lambda}_L$, decays as the boundary deformation parameter $\phi_0/T$ increases, which is consistent with the Lyapunov exponent vanishing at the zero temperature limit (see, for example, \cite{Prihadi2023}). Moreover, the decay process is slowed down by the presence of angular momentum, in agreement with the fact that it increases the surface gravity of the black hole. Additionally, the relation between Lyapunov exponent $\bar{\lambda}_L$ and the interior geometry of the black hole is illustrated in the plots of $\bar{\lambda}_L$ versus the Kasner exponent $p_t$.\\
\indent While the Lyapunov exponent is sensitive to variations in angular momentum $\mathcal{L}$, it remains largely unaffected by changes in the charge density $\rho$. As shown in \cite{Malvimat2022,Malvimat2023,Malvimat2023b,Prihadi2023}, the chaos bound of the black hole is modified due to rotation. However, how it is modified due to both rotation and charge (as another conserved quantity) remains unclear. In \cite{halder2019globalsymmetrymaximalchaos}, it is shown that the upper bound to the Lyapunov exponent can be loosened by adding a chemical potential $\mu$. For the rotating case, the chemical potential is the angular velocity $\Omega_H$ and the extra term in the upper bound gets corrected by a factor $1-\Omega_H\mathcal{L}$. Our numerical results show that the chaotic properties of the black hole are significantly influenced by the rotation parameters, while they are not substantially affected by the electric charge.\\
\indent On the other hand, a completely different behavior emerges in the scrambling time delay $\Delta\tau$. Since the delay arises from the charge interaction between the black hole and the shock waves, $\Delta\tau$ heavily depends on the charge density $\rho$. The scrambling time delay $\Delta\tau$ grows with increasing charge density $\rho$ and reduces to zero when the black hole is neutral, even if it is rotating. The angular momentum $\mathcal{L}$ also influences $\Delta\tau$, which grows as $\mathcal{L}$ increases. Similarly, the rotation parameter $N_{H1}$ of the black hole increases $\Delta\tau$. In general, $\phi_0/T$ reduces the scrambling time delay due to the trade-off relation between $\phi$ and $\mu_e$. We expect that an Einstein-Maxwell-Scalar (EMS) coupling (such as in \cite{Sword2022a,prihadi2025scramblingchargedhairyblack}), could make the trade-off relation even stronger.\\
\indent The effects of rotation and charge to the chaos bound has been previously studied in \cite{Prihadi2023} for Kerr–Sen–AdS black holes, where it was found that the Lyapunov exponent can violate the rotating chaos bound proposed in \cite{Malvimat2023} for finely tuned cases with dilaton and axion charges, although in general the bound is respected. In this work, we focus instead on the effect of the boundary scalar deformation $\phi_0$ on the Lyapunov exponent. As shown in \cite{prihadi2025scramblingchargedhairyblack}, such deformations decrease the ratio of $\lambda_L$ to the surface gravity $\kappa$, suggesting that turning on $\phi_0$ does not lead to a violation of the chaos bound. Consistently, we find that $\phi_0/T$ lowers the normalized Lyapunov exponent $\bar{\lambda}_L$, which can be approximately interpreted as the ratio $\lambda_L/\kappa$ to some extent (see for instance, the case when the black hole approach extremality \cite{Prihadi2023}). Our results further show that while rotation parameters ($\mathcal{L}$ and $N_{H1}$) enhance the normalized $\lambda_L$, it never exceeds unity, and thus the chaos bound remains respected in the presence of scalar boundary deformations.\\
\indent In this work, we primarily focus on the first Kasner regime, as the study of the Lyapunov exponent requires access only to the interior region not far from the horizon. A more comprehensive exploration of the deep interior, both analytical and numerical, is left for future work. It is plausible that novel features, such as Kasner inversions or transitions, could emerge at large values of the radial coordinate $r$. In particular, \cite{Sword2022a} shows that the Einstein-Maxwell-Scalar (EMS) coupling parameter can drive Kasner transitions, while \cite{Gao2024} demonstrates that rotation may induce Kasner inversion, which potentially leads to a more stable Kasner regime. In this case, we expect two types of transitions, one associated with $N(r)$ and the other with $A_t(r)$. One of them is expected to correspond to the transition from $c^2 < 1$ to $c^2 > 1$, while the other may give rise to a change from $c^2 < 1/3$ to $c^2 > 3$. Another intriguing direction involves the study of scalar field oscillations within the black hole interior, particularly in the context of holographic superconductors \cite{Hartnoll2021, Gao2024}. We leave the development of analytical expressions for the ER bridge collapse, scalar field oscillations, and Kasner inversion/transitions in this black hole background to future work.
\section*{Acknowledgement}
We would like to thank Li Li for useful comments on the draft and bringing Ref. \cite{An2021} to us. H. L. P. would like to thank the National Research and Innovation Agency (BRIN) for its financial support through the Postdoctoral Program. R. R. F. would like to thank the Ministry of Education and Culture (Kemendikbud) Republic of Indonesia for financial support through 
BPI Kemendikbudristek Scholarship. F. K. would like to thank the Ministry of Education and Culture (Kemendikbud) Republic of Indonesia for financial support through Beasiswa Unggulan Scholarship. D. D. is supported by the APCTP (YST program) through the Science and Technology Promotion Fund and Lottery Fund of the Korean Government and the Korean Local governments, Gyeongsangbuk-do Province, and Pohang city. F. P. Z. would like to thank the Ministry of Higher Education, Science, and Technology (Kemendiktisaintek) Republic of Indonesia for financial support.  H. L. P. and F. K. would like to thank members of Research Center for Quantum Physics, National Research and Innovation Agency, and Theoretical Physics Group, Institut Teknologi Bandung, for valuable discussions.
\section*{Appendix A: Near-Horizon Expansion Coefficients}
The near-horizon expansion coefficients in 
Eqs. \eqref{nearhorizonphi}-\eqref{nearhorizonn} are given by
\begin{align}
    \phi_{H1}=-\frac{2 A_{xH1}^3 \text{$\phi_{H0}$}}{N_{H1}\left(\sqrt{-A_{xH1}^2 e^{\text{$\chi_{H0}$}} \left(A_{xH1}^2 \left(N_{H1}^2 e^{\text{$\chi_{H0}$}}-4 \left(\text{$\phi_{H0}^2$}+3\right)\right)-4N_{H1}^2 e^{\text{$\chi_{H0}$}}\right)}+2 A_{xH1}N_{H1} e^{\text{$\chi_{H0}$}}\right)}
\end{align}
\begin{align}
    f_{H1}=\frac{2N_{H1}^2 e^{\text{$\chi_{H0}$}}}{A_{xH1}^2}+\frac{N_{H1}\sqrt{-A_{xH1}^2 e^{\text{$\chi_{H0}$}} \left(A_{xH1}^2 N_{H1}^2 e^{\text{$\chi_{H0}$}}-4A_{xH1}^2 \text{$\phi_{H0}$}^2-12A_{xH1}^2-4N_{H1}^2 e^{\text{$\chi_{H0}$}}\right)}}{A_{xH1}^3}.
\end{align}
\begin{align}
    \chi_{H0}=\frac{2 e^{-\text{$\chi_{H0}$}} (\sqrt{-A_{xH1}^2 e^{\text{$\chi_{H0}$}} (A_{xH1}^2 (N_{H1}^2 e^{\text{$\chi_{H0}$}}-4 (\text{$\phi_{H0}$}^2+3))-4 N_{H1}^2 e^{\text{$\chi_{H0}$}})}+4A_{xH1}N_{H1} e^{\text{$\chi_{H0}$}})}{A_{xH1}N_{H1}}
\end{align}
\begin{align}
    N_{H0}=&\frac{A_{tH1}A_{xH1}+2N_{H1}^2}{A_{xH1}^2}+\\\nonumber
    &+\frac{e^{-\text{$\chi_{H0}$}} \sqrt{-A_{xH1}^4N_{H1}^2 e^{2 \text{$\chi_{H0}$}}+12 A_{xH1}^4 e^{\text{$\chi_{H0}$}}+4A_{xH1}^4 e^{\text{$\chi_{H0}$}} \text{$\phi_{H0}$}^2+4 A_{xH1}^2 N_{H1}^2 e^{2 \text{$\chi_{H0}$}}}}{A_{xH1}^3}
\end{align}
\section*{Appendix B: Numerical Solutions Without Scalar Field}
In this appendix, we provide the numerical solution without scalar field, i.e. when $\phi=0$. Note that, in contrast with the case when $A_x=0$, in this case, $\chi$ does not goes to 0 when $\phi\rightarrow0$. In this work, the analytical solution to the case without scalar field is not considered. 
\begin{figure}
\includegraphics[scale=0.6]{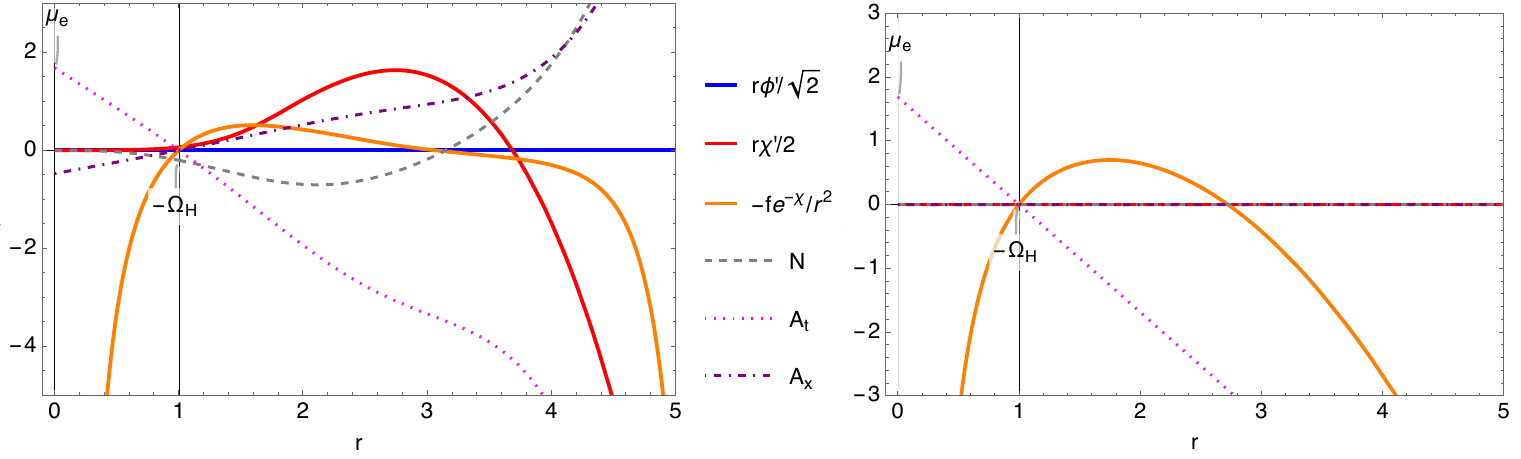}
\caption{\label{fig:noscalar}Numerical solutions when $\phi=0$. Left: the solution with $A_x\neq0$ and $N\neq0$. Right: the solution when both $A_x$ and $N$ vanish.}
\end{figure}
Numerical solution without scalar field can be seen in Figure \ref{fig:noscalar}. Here, one can see that both solutions (with or without $A_x$ and $N$) have inner horizon.
\section*{Appendix C: Remarks on the No Inner Horizon}
Observe that Eq. \eqref{EoMphi} can be written as
\begin{align}
    \left( \frac{f  e^{-\chi/2}\phi'}{r^2} \right)' &= \frac{f e^{-\chi/2} \phi}{r^2} \bigg(\frac{2 q^2 A_tA_xNe^{\chi}}{f^2}-\frac{q^2A_t^2 e^{\chi}}{f^2}-\frac{q^2A_x^2N^2 e^{\chi}}{f^2}+\frac{q^2A_x^2}{f}+\frac{m^2}{r^2 f}\bigg)  \\ &= \frac{f e^{-\chi/2} \phi}{r^2} \bigg(-\frac{q^{2}e^{\chi}}{f^2}(A_{t} - A_x N)^{2}+\frac{q^2A_x^2}{f}+\frac{m^2}{r^2 f}\bigg)\label{eq:noinnerhorizon}
\end{align}
where we have used Eq. \eqref{EoMEins3} in the last line. If there were two horizons $f(r_H) = f(r_I) = 0$ with outer horizon $r_H$ and inner horizon $r_I$, from Eq. \eqref{eq:noinnerhorizon} we would have
\begin{align}
    0 &= \int_{r_H}^{r_I} \left(\frac{f e^{-\chi/2}\phi\phi'}{r^2}\right)' dr = \int_{r_H}^{r_I} \frac{e^{-\chi/2}}{r^2} \Bigg[ f\phi'^2 + \phi^2 \bigg(-\frac{q^{2}e^{\chi}}{f}(A_{t} - A_x N)^{2}+q^2A_x^2+\frac{m^2}{r^2}\bigg) \Bigg] dr .
\end{align}
The first equality was obtained because $f(r_H) = f(r_I) = 0$. The function $f(r)$ is always negative between $r_H$ and $r_I$. When $m^2<0$, one cannot conclude that the term in the right parenthesis is always negative (or positive) due to the remaining terms which is always positive in this interval. Hence, one cannot conclude that $f(r_I)=0$ cannot exist using this method.
\bibliographystyle{elsarticle-num}
\bibliography{BIBInterior.bib}

\end{document}